# Analytical transit light curves for arbitrary power-law limb darkening: a unified framework

F. A. Chishtie,[1,2★] M. I. Saeed[1] and S. N. Goderya[3]

[1]*Peaceful Society Science and Innovation Foundation, Vancouver, BC V6K 2E8, Canada*
[2]*Department of Occupational Science and Occupational Therapy, University of British Columbia, Vancouver, BC V6T 2B5, Canada*
[3]*Department of Physics and Astronomy, Tarleton State University, Stephenville, TX 76401, USA*

## ABSTRACT

We present a unified analytical framework for computing exoplanet transit light curves under arbitrary real power-law limb darkening $I(\mu) = I_0\mu^\alpha$, $\alpha > -2$, eliminating the two-decade restriction to integer-polynomial forms. Our central result is an exact closed-form expression for the normalized stellar flux in terms of Appell's bivariate hypergeometric function $F_1$, valid for all real $\alpha$ and all geometric transit configurations. Three mutually equivalent formulations – a geometric kernel representation, a Riemann–Liouville fractional-calculus framework, and a hypergeometric-integrand representation – provide complementary physical insights and independent computational routes, with inter-method agreement at $|\Delta\mathcal{F}| \lesssim 10^{-17}$. The framework encompasses the square-root law ($\alpha = 1/2$) essential for M-dwarf characterization, half-integer powers required by Claret's four-parameter law, and arbitrary real values enabling empirical fitting, while recovering integer-polynomial elliptic-integral expressions as special cases. Validation at 40-digit precision across five geometric regions and seven $\alpha$ values confirms a $\sim 10^4\times$ computational speedup over Monte Carlo integration and thirteen orders of magnitude accuracy advantage. Linear superposition enables exact treatment of arbitrary multiparameter limb-darkening laws without special-case logic.

## 1 INTRODUCTION

The accurate characterization of transiting exoplanets depends critically on precise modelling of stellar limb darkening. As photometric facilities achieve sub-ppm precision – *Kepler* (W. J. Borucki et al. 2010), *TESS* (G. R. Ricker et al. 2015), and *James Webb Space Telescope* (*JWST;* J. P. Gardner et al. 2006; Z. Rustamkulov et al. 2023; L. Alderson et al. 2023) – systematic errors from inadequate limb-darkening models now dominate the error budget for radius and atmospheric composition measurements.

### 1.1 The polynomial approximation problem

The standard approach, originating in the eclipse modelling framework of Kopal (1950) and formalized for transits by K. Mandel & E. Agol (2002) represents the limb-darkening profile as a polynomial in $\mu = \sqrt{1-\rho^2}$ (where $\rho$ is the normalized projected stellar radius):

$$\frac{I(\mu)}{I(1)} = \sum_{k=0}^{N} c_k \mu^k. \tag{1}$$

Restricting to integer exponents permits exact analytical integration through elliptic integrals, but forces all physically motivated profiles to be approximated. The four-parameter law (A. Claret 2000),

$$\frac{I(\mu)}{I(1)} = 1 - \sum_{k=1}^{4} a_k(1-\mu^{k/2}), \tag{2}$$

requires half-integer powers $\mu^{1/2}$, $\mu^{3/2}$ inaccessible to the integer framework. For cool M-dwarf atmospheres ($T_{\rm eff} \lesssim 3500$ K), the square-root law $I(\mu) \propto \mu^{1/2}$ is physically preferred (J. Díaz-Cordovés & A. Giménez 1992; A. Claret, P. H. Hauschildt & S. Witte 2012), yet cannot be treated exactly within integer-polynomial frameworks. The power-2 law (D. Hestroffer 1997) and logarithmic law (D. A. Klinglesmith & S. Sobieski 1970) similarly resist exact polynomial representation, introducing systematic errors at the $\sim 10^{-3}$ level in relative flux (D. K. Sing 2010; Espinoza & Jordan 2015), now comparable to or exceeding statistical uncertainties in *JWST* observations.

### 1.2 Non-integer exponents and physical motivation

Stellar atmosphere calculations (Z. Magic et al. 2015; A. Claret & S. Bloemen 2011) consistently produce limb-darkening profiles better described by non-integer power laws. The power-law form $I(\mu) = I_0\mu^\alpha$ with real $\alpha$ arises naturally from grey atmosphere theory and provides a one-parameter family interpolating between uniform ($\alpha = 0$) and strongly darkened ($\alpha \gg 1$) profiles.

★ E-mail: fachisht@uwo.ca

Allowing $\alpha > -2$ (which includes limb-brightening for $-2 < \alpha < 0$, relevant for chromospheric emission) provides the most general single-power-law family. The condition $\alpha > -2$ is precisely the requirement for finite total stellar flux, as follows from $\int_0^1 (1-\rho^2)^{\alpha/2}\rho\,\mathrm{d}\rho = 1/(\alpha+2)$.

### 1.3 Prior work on non-integer limb darkening

A. Pál (2008) derived exact analytical transit flux formulae for the square-root law using direct integration in terms of complete elliptic integrals $K(k)$, $E(k)$, and $\Pi(n,k)$. A. Pál (2012) extended this to the three-halves law. These case-by-case derivations demonstrate the feasibility of exact treatment but do not scale to a general-$\alpha$ framework. A. Giménez (2006) expressed transit light curves in terms of Jacobi polynomials but maintained the polynomial restriction. D. Hestroffer (1997) used power-law limb darkening for stellar disc integration but did not derive transit light curve formulae. Building on the analytical Green's-theorem framework of Luger et al. (2019) for spherical-harmonic surface maps, the systematic work of E. Agol, R. Luger & D. Foreman-Mackey (2020) on polynomial limb darkening provides an efficient $\mathcal{O}(1)$ recursion for integer cases. No existing analytical framework handles arbitrary real $\alpha$, leaving a fundamental gap in our ability to model physically motivated limb darkening with the precision demanded by modern observations.

### 1.4 Our contribution

We derive exact closed-form transit light curves for *arbitrary real*$\alpha > -2$, expressed in terms of Appell's bivariate hypergeometric function $F_1$ (P. Appell 1880). The key mathematical tools are: (i) a geometric kernel representation of the obscured stellar flux; (ii) integration by parts exploiting the boundary structure of the transit geometry; and (iii) the Euler integral representation of $F_1$ matched to the resulting one-dimensional (1D) integral. The Appell $F_1$ function is an analytic result in precisely the same sense as the elliptic integrals of K. Mandel & E. Agol (2002): it is a well-characterized special function implemented in all major numerical libraries (`mpmath.appellf1`, Mathematica's `AppellF1`). No two-dimensional (2D) integration over the stellar disc is required. For integer $\alpha$, the Appell series truncates finitely, recovering established elliptic-integral expressions exactly.

### 1.5 Superposition and multiparameter laws

For a general limb-darkening profile expressible as a superposition of power laws, $I(\mu)/I(1) = \sum_j c_j \mu^{\alpha_j}$, the transit flux is

$$\mathcal{F} = \frac{\sum_j c_j F_{\mathrm{vis}}(p,z,\alpha_j)}{\sum_j c_j/(\alpha_j+2)}, \tag{3}$$

reducing the multiparameter problem to a sum of independent power-law evaluations without special-case logic. Claret's four-parameter law thus reduces to five power-law evaluations at $\alpha_j \in \{0, 1/2, 1, 3/2, 2\}$, all covered by our framework.

### 1.6 Paper structure

Section 2 establishes the geometric kernel representation and its measure-theoretic foundations. Section 3 develops the fractional-calculus framework, derives the Appell $F_1$ closed forms through an explicit six-step calculation, and establishes the $\alpha$-differentiation identity and continuous recursion relations. Section 4 presents the hypergeometric-integrand representation. Section 5 covers half-integer and multiparameter special cases. Section 6 provides comprehensive analytical and numerical validation. Section 7 discusses astrophysical implications and future directions, and Section 8 concludes.

## 2 GEOMETRIC FRAMEWORK

### 2.1 Coordinate systems and intensity profile

We adopt normalized coordinates with stellar radius $R_\star = 1$. The planet has radius ratio $p = R_p/R_\star$ with centre at normalized separation $z = d/R_\star$ from the stellar centre. Polar coordinates $(\rho,\theta)$ on the stellar disc have $\rho \in [0,1]$ as the normalized radial coordinate. The projection cosine from disc centre relates to $\rho$ by $\mu = \sqrt{1-\rho^2}$, and the power-law intensity profile is

$$I(\rho) = I_0\,(1-\rho^2)^{\alpha/2}, \qquad 0 \le \rho \le 1, \quad \alpha > -2, \tag{4}$$

where $I_0 > 0$ is the central intensity.

### 2.2 Total flux and normalization

The total unobscured stellar flux integrates the intensity over the disc:

$$F_{\mathrm{tot}}(\alpha) = 2\pi I_0 \int_0^1 (1-\rho^2)^{\alpha/2}\rho\,\mathrm{d}\rho = \frac{2\pi I_0}{\alpha+2}, \tag{5}$$

where the substitution $u = 1-\rho^2$ yields $\int_0^1 (1-\rho^2)^{\alpha/2}\rho\,\mathrm{d}\rho = 1/(\alpha+2)$, which converges for all $\alpha > -2$. The normalized flux is $\mathcal{F}(p,z,\alpha) := F_{\mathrm{vis}}/F_{\mathrm{tot}}$ and the blocked fraction is $B(p,z,\alpha) := 1-\mathcal{F}$. Throughout, $\mathcal{D}$ denotes the lens-shaped overlap region $\{(\rho,\phi) \in \mathbb{R}^2 : \rho \le 1,\ |\boldsymbol{\rho} - \mathbf{z}| \le p\}$.

### 2.3 Geometric regions

The transit geometry divides $(p,z)$ parameter space into five regions. Region 1 ($z \ge 1+p$) has no overlap and $\mathcal{F} = 1$. Region 2 ($|1-p| < z < 1+p$) is partial eclipse where $\mathcal{D}$ is a lens (bigon). Region 3a ($z < 1-p$, $p \le 1$) has the planet fully interior to the stellar disc. Region 3b ($z < p-1$, $p > 1$) has the star fully within the planet, so $\mathcal{F} = 0$. Region 3c ($p-1 < z < p+1$, $p > 1$) is the large-planet partial eclipse, analogous to Region 2 with modified radial limits.

### 2.4 Angular integration and the kernel representation

After integrating out the azimuthal angle, the blocked-flux integral reduces to

$$B(p,z,\alpha) = \frac{\alpha+2}{\pi} \int_{r_1}^{r_2} \arccos\left(\frac{\rho^2+z^2-p^2}{2\rho z}\right)(1-\rho^2)^{\alpha/2}\rho\,\mathrm{d}\rho, \tag{6}$$

where $r_1 = |p-z|$ and $r_2 = \min(1, p+z)$, and the factor $(\alpha+2)/\pi$ follows from the normalization equation (5) together with the symmetry of the lens under reflection through the star–planet axis. The kernel function $A(\rho,z,p) := (\rho^2+z^2-p^2)/(2\rho z)$ is the cosine of the half-subtended angle at radial distance $\rho$.

### 2.5 Cap–lens decomposition

The integral $B$ decomposes as $B = B_{\rm cap} + B_{\rm lens}$, where

$$B_{\rm cap}(p, z, \alpha) = \mathbf{1}_{(p>z)}\left[1 - (1 - (p - z)^2)^{1+\alpha/2}\right], \tag{7}$$

is evaluated exactly via $u = 1 - \rho^2$ over the fully-blocked annulus $\rho \in [0, |p - z|]$, and the lens integral

$$B_{\rm lens}(p, z, \alpha) = \frac{\alpha + 2}{\pi} \int_{r_1}^{r_2} \arccos\left(\frac{\rho^2 + z^2 - p^2}{2\rho z}\right) \times (1 - \rho^2)^{\alpha/2} \rho \, \mathrm{d}\rho \tag{8}$$

carries the geometrically non-trivial contribution requiring the Appell $F_1$ treatment.

### 2.6 Regularity and measure-theoretic foundations

For fixed $\alpha > -2$, the normalized flux $\mathcal{F}(p, z, \alpha)$ is continuous in $(p, z) \in \mathbb{R}^+ \times \mathbb{R}^+$. The integrand $\arccos A(\rho) \cdot (1 - \rho^2)^{\alpha/2} \rho$ lies in $L^1([0, 1])$, with a dominating function $(1 - \rho^2)^{\alpha/2} \rho$ that is independent of $(p, z)$ for each fixed $\alpha > -2$. By the Lebesgue dominated convergence theorem (Rudin, 1987), $\lim_{(p',z')\to(p,z)} B(p', z', \alpha) = B(p, z, \alpha)$, establishing continuity. As $z \to 1 + p$, the lens interval $[r_1, r_2]$ collapses to measure zero, giving $B_{\rm lens} \to 0$ and $\mathcal{F} \to 1$, matching Region 1. At all region boundaries, continuity of $B$ follows similarly from the collapse of the arccos kernel to its boundary values 0 or $\pi$.

### 2.7 Central transit closed form

For $z = 0$, $p < 1$, the integration angle spans the full circle at every radius $\rho < p$, giving

$$B(p, 0, \alpha) = (\alpha + 2) \int_0^p (1 - \rho^2)^{\alpha/2} \rho \, \mathrm{d}\rho = 1 - (1 - p^2)^{1+\alpha/2}, \tag{9}$$

so $\mathcal{F}(p, 0, \alpha) = (1 - p^2)^{1+\alpha/2}$ exactly, requiring no numerical integration. For $\alpha = 0$ (uniform disc) this recovers $\mathcal{F} = 1 - p^2$.

## 3 FRACTIONAL-CALCULUS FRAMEWORK: EXACT APPELL $F_1$ CLOSED FORMS

We now derive exact closed-form expressions for $B_{\rm lens}$ valid for arbitrary real $\alpha > -2$, using Riemann–Liouville fractional calculus to reduce the 1D integral (8) to Appell bivariate hypergeometric functions. The derivation proceeds through six explicit steps.

### 3.1 Riemann–Liouville fractional calculus

DEFINITION 3.1 (Riemann–Liouville fractional integral). For $\sigma > 0$ and integrable $f$ on $[0, t]$:

$$(I^\sigma f)(t) = \frac{1}{\Gamma(\sigma)} \int_0^t (t - \tau)^{\sigma-1} f(\tau) \, \mathrm{d}\tau. \tag{10}$$

The fundamental power-law formula (S. G. Samko, A. A. Kilbas & O. I. Marichev 1993; A. A. Kilbas, H. M. Srivastava & J. J. Trujillo 2006; K. B. Oldham & J. Spanier 1974; Podlubny, 1999) is

$$I^\alpha t^\gamma = \frac{\Gamma(\gamma + 1)}{\Gamma(\gamma + \alpha + 1)} t^{\gamma+\alpha}, \qquad \gamma > -1, \quad \alpha > 0. \tag{11}$$

This formula enables exact evaluation of $(1 - \rho^2)^{\alpha/2}$ integrals for arbitrary real $\alpha$ without polynomial approximation.

### 3.2 Boundary behaviour

LEMMA 3.2 (Boundary behaviour of the IBP product). *For $\alpha > -2$ and $(p, z)$ with $z > 0$, the function $(1 - \rho^2)^{1+\alpha/2} \arccos A(\rho)$ satisfies*

$$\left[(1 - \rho^2)^{1+\alpha/2} \arccos A(\rho)\right]_{\rho=r_1}^{\rho=r_2} = -\mathbf{1}_{(p>z)} (1 - r_1^2)^{1+\alpha/2}, \tag{12}$$

*where $\mathbf{1}_{(p>z)}$ is the indicator of the subcase $p > z$.*

*Proof.* At $\rho = r_2 = \min(1, p + z)$: $A(r_2) = 1$, hence $\arccos A(r_2) = 0$ and the product vanishes. At $\rho = r_1 = |p - z|$ with $r_1 > 0$: $A(r_1) = +1$ when $p < z$, giving $\arccos A(r_1) = 0$; while $A(r_1) = -1$ when $p > z$, giving $\arccos A(r_1) = \pi$. The signed difference of endpoint values then yields the stated result. □

### 3.3 Integration-by-parts reduction

*For $z > 0$:*

$$B_{\rm lens}(p, z, \alpha) = \mathbf{1}_{(p>z)} (1 - r_1^2)^{1+\alpha/2} - \frac{1}{\pi} \int_{r_1}^{r_2} \frac{(\rho^2 + p^2 - z^2)(1 - \rho^2)^{1+\alpha/2}}{\rho \sqrt{\Delta(\rho)}} \, \mathrm{d}\rho, \tag{13}$$

*where $\Delta(\rho) = 4\rho^2 z^2 - (\rho^2 + z^2 - p^2)^2$.*

*Proof.* Integrate equation (8) by parts with $u = \arccos A(\rho)$ and $dv = (\alpha + 2)(1 - \rho^2)^{\alpha/2} \rho \, d\rho$, so that $v = -(1 - \rho^2)^{1+\alpha/2}$. Computing $du/d\rho = -A'(\rho)/\sqrt{1 - A(\rho)^2}$ with

$$A'(\rho) = \frac{\rho^2 + p^2 - z^2}{2\rho^2 z}, \qquad \sqrt{1 - A(\rho)^2} = \frac{\sqrt{\Delta(\rho)}}{2\rho z}, \tag{14}$$

yields $d(\arccos A)/d\rho = -(\rho^2 + p^2 - z^2)/(\rho\sqrt{\Delta})$. By Lemma 3.2 the boundary term contributes $-\mathbf{1}_{(p>z)} (1 - r_1^2)^{1+\alpha/2}$ to $\int u \, dv$, and the integration-by-parts formula gives equation (13). □

### 3.4 Angular mean of a power function

A key identity is needed for Region 3a, connecting angular averages of power functions to hypergeometric functions.

LEMMA 3.4 (Angular mean of a power function on a circle). *Let $a > b \geq 0$ and $\lambda > -1$. Then*

$$\frac{1}{2\pi} \int_0^{2\pi} (a - b\cos\phi)^\lambda \, \mathrm{d}\phi = a^\lambda \, {}_2F_1\left(-\lambda, \tfrac{1}{2}; 1; \frac{b^2}{a^2}\right). \tag{15}$$

*Proof.* Write $(a - b\cos\phi)^\lambda = a^\lambda (1 - (b/a)\cos\phi)^\lambda$ and expand via the binomial series. Integrating term by term and using $\frac{1}{2\pi}\int_0^{2\pi} \cos^{2m}\phi \, \mathrm{d}\phi = (1/2)_m/m!$ with odd moments vanishing, only even terms survive. The resulting sum matches the hypergeometric series ${}_2F_1(-\lambda, 1/2; 1; b^2/a^2)$ term by term after applying the Legendre duplication formula. □

### 3.5 Appell bivariate hypergeometric function

The Appell bivariate hypergeometric function (P. Appell 1880; G. E. Andrews, R. Askey & R. Roy 1999) is defined by its Euler integral representation,

$$F_1(a;\, b_1, b_2;\, c;\, x, y) = \frac{\Gamma(c)}{\Gamma(a)\Gamma(c-a)} \int_0^1 t^{a-1}(1-t)^{c-a-1} \times (1-xt)^{-b_1}(1-yt)^{-b_2}\, \mathrm{d}t, \qquad (16)$$

valid for $c > a > 0$ and $(1-xt), (1-yt) \neq 0$ on [0,1]; analytically continued otherwise. For integer values of $b_1$ or $b_2$ the Pochhammer symbols $(b_j)_n$ in the double series representation truncate finitely, reducing $F_1$ to complete elliptic integrals.

### 3.6 Exact Appell $F_1$ closed forms: six-step derivation

We derive the closed forms through six explicit steps for Region 3a; the Region 2 derivation is analogous and stated thereafter.

#### 3.6.1 Step 1: discriminant factorization

The discriminant is $\Delta(\rho) = ((p+z)^2 - \rho^2)(\rho^2 - r_1^2)$ with $r_1 = |p-z|$. Setting $r_2' = p+z$ (the geometric limit regardless of whether $r_2'$ exceeds 1), we write $\Delta = r_2'^2(\rho^2 - r_1^2)(1 - \rho^2/r_2'^2)$ so that $\sqrt{\Delta} = \sqrt{(r_2'^2 - \rho^2)(\rho^2 - r_1^2)}$.

*Step 2: substitution for region 3a ($r_2 = p+z \leq 1$).*

Set $t = (\rho^2 - r_1^2)/(4pz)$, so $t \in [0,1]$ as $\rho$ ranges over $[r_1, r_2]$. The discriminant factorizes as $\sqrt{\Delta} = 4pz\sqrt{t(1-t)}$, and

$$1 - \rho^2 = (1 - r_1^2)(1 - nt), \qquad n := \frac{4pz}{1 - r_1^2} < 1. \qquad (17)$$

The numerator of equation (13) evaluates as $\rho^2 + p^2 - z^2 = 2p(p - z + 2zt)$, using $r_1^2 = (p-z)^2$. After the change of variable $\mathrm{d}\rho = 2pz\,\mathrm{d}t/\rho$, the integral becomes

$$B_{\rm lens}^{(3a)} = \mathbf{1}_{(p>z)}\,(1 - r_1^2)^{1+\alpha/2} - \frac{p\,(1-r_1^2)^{1+\alpha/2}}{\pi\, r_1^2} \int_0^1 \frac{(p - z + 2zt)\,(1-nt)^{1+\alpha/2}}{(1+\lambda t)\sqrt{t(1-t)}}\,\mathrm{d}t, \qquad (18)$$

where $\lambda := 4pz/r_1^2$.

*Step 3: splitting the linear factor.*

Write $(p - z + 2zt) = (p-z)\cdot 1 + 2z\cdot t$ and decompose accordingly:

$$B_{\rm lens}^{(3a)} = \mathbf{1}_{(p>z)}\,(1 - r_1^2)^{1+\alpha/2} - \frac{p\,(1-r_1^2)^{1+\alpha/2}}{\pi\, r_1^2}\Bigg[(p-z)\int_0^1 \frac{(1-nt)^{1+\alpha/2}}{(1+\lambda t)\sqrt{t(1-t)}}\,\mathrm{d}t + 2z\int_0^1 \frac{t\,(1-nt)^{1+\alpha/2}}{(1+\lambda t)\sqrt{t(1-t)}}\,\mathrm{d}t\Bigg]. \qquad (19)$$

#### 3.6.2 Step 4: matching to the euler integral (16)

For the first integral, set $a = \frac{1}{2}$, $c = 1$, so $\Gamma(c)/[\Gamma(a)\Gamma(c-a)] = 1/\pi$, $b_1 = -(1+\alpha/2)$, $b_2 = -1$ (from the $(1+\lambda t)^{-1}$ factor), $x = n$, $y = -\lambda$:

$$\int_0^1 \frac{(1-nt)^{1+\alpha/2}}{(1+\lambda t)\sqrt{t(1-t)}}\,\mathrm{d}t = \pi\, F_1\big(\tfrac{1}{2};\, -\nu,\, 1;\, 1;\, n,\, -\lambda\big), \qquad \nu = 1 + \tfrac{\alpha}{2}. \qquad (20)$$

For the second integral, set $a = \frac{3}{2}$, $c = 2$, giving $\Gamma(2)/[\Gamma(3/2)\Gamma(1/2)] = 2/\pi$:

$$\int_0^1 \frac{t\,(1-nt)^{1+\alpha/2}}{(1+\lambda t)\sqrt{t(1-t)}}\,\mathrm{d}t = \frac{\pi}{2}\, F_1\big(\tfrac{3}{2};\, -\nu,\, 1;\, 2;\, n,\, -\lambda\big). \qquad (21)$$

*Step 5: assembly for region 3a*

Collecting equations ()–(21):

$$B_{\rm lens}^{(3a)} = \mathbf{1}_{(p>z)}\,(1 - r_1^2)^{\nu} - \frac{p\,(1-r_1^2)^{\nu}}{\pi\, r_1^2}\Big[\pi(p-z)\,F_1\big(\tfrac{1}{2};\, -\nu, 1; 1; n, -\lambda\big) + \pi z\, F_1\big(\tfrac{3}{2};\, -\nu, 1; 2; n, -\lambda\big)\Big], \qquad (22)$$

which simplifies (the $\pi$ factors cancel) to the stated theorem below.

#### 3.6.3 Step 6: region 2 analogously

For $p+z > 1$, set $t = (\rho^2 - r_1^2)/(1 - r_1^2)$ with $t \in [0,1]$ as $\rho$ ranges over $[r_1, 1]$. The discriminant now gives $\sqrt{\Delta} = (1 - r_1^2)\sqrt{n}\sqrt{t(1-t/n)}$ with $n = 4pz/(1-r_1^2) > 1$. Setting $s = t/n$ converts to the Euler form with argument $1/n < 1$, and the second factor $(1-\rho^2)^{1+\alpha/2}$ provides the $(1 - s\cdot(1-r_1^2))$ term that defines the parameter $\lambda'$. The resulting integrals match equation (16) with $a = \frac{1}{2}$ and $a = \frac{3}{2}$ as before, with normalization constants $C_1 = \sqrt{\pi}\,\Gamma(\nu+1)/\Gamma(\nu+\frac{3}{2})$ and $C_2 = \sqrt{\pi}\,\Gamma(\nu+1)/[2\Gamma(\nu+\frac{5}{2})]$ arising from the additional $\Gamma$ factors associated with the power $\nu$.

We now state the two main theorems.

THEOREM 3.5 (Exact Appell $F_1$ closed form, Region 3a). *For $p+z \leq 1$, $\alpha > -2$, $z > 0$, with $\nu = 1 + \alpha/2$, $r_1 = |p-z|$, $n = 4pz/(1-r_1^2) < 1$, and $\lambda = 4pz/r_1^2$:*

$$B_{\rm lens}^{(3a)} = \mathbf{1}_{(p>z)}\,(1 - r_1^2)^{\nu} - \frac{p\,(1-r_1^2)^{\nu}}{r_1^2}\Big[(p-z)\,F_1\big(\tfrac{1}{2};\, -\nu, 1; 1; n, -\lambda\big) + z\, F_1\big(\tfrac{3}{2};\, -\nu, 1; 2; n, -\lambda\big)\Big]. \qquad (23)$$

*The Euler integral (16) converges absolutely since $(1-nt) > 0$ and $(1+\lambda t) > 0$ for all $t \in [0,1]$.*

REMARK 3.6 (Sign and subcases of $B_{\rm lens}^{(3a)}$). When $p < z$, the indicator term vanishes, $(p-z) < 0$, and the bracketed combination of Appell $F_1$ functions is negative, so the second term of equation (23) is positive, yielding $B_{\rm lens}^{(3a)} > 0$. When $p > z$, the indicator term contributes the full annular block $(1-r_1^2)^{\nu}$, from which the partial-coverage correction is subtracted; the cap term $B_{\rm cap} = 1 - (1-r_1^2)^{\nu}$ in equation (7) accounts separately for the inner disc $\rho \in [0, r_1]$ where the planet covers the full angular range. Convergence of the Euler representation for $n < 1$ and $\lambda > 0$ is guaranteed by the denominators remaining bounded away from zero on [0,1].

THEOREM 3.7 (Exact Appell $F_1$ closed form, Region 2). *For $p+z>1$, $\alpha>-2$, with $\nu=1+\alpha/2$, $r_1=|p-z|$, $n=4pz/(1-r_1^2)>1$, $\lambda'=(1-r_1^2)/r_1^2$, $C_1=\sqrt{\pi}\,\Gamma(\nu+1)/\Gamma(\nu+\frac{3}{2})$, $C_2=\sqrt{\pi}\,\Gamma(\nu+1)/[2\,\Gamma(\nu+\frac{5}{2})]$:*

$$B_{\rm lens}^{(2)}=\frac{(1-r_1^2)^{\nu}}{2\pi\sqrt{n}\,r_1^2}\Big[\quad 2pr_1\,C_1\,F_1\big(\tfrac{1}{2};\ \tfrac{1}{2},1;\ \nu+\tfrac{3}{2};\ \tfrac{1}{n},-\lambda'\big) - (1-r_1^2)C_2\,F_1\big(\tfrac{3}{2};\ \tfrac{1}{2},1;\ \nu+\tfrac{5}{2};\ \tfrac{1}{n},-\lambda'\big)\Big]. \quad (24)$$

*For Region 2 with $\lambda'>1$, the Appell double series diverges but the Euler integral (16) converges absolutely since the denominator factors $(1-t/n)$ and $(1+\lambda' t)$ remain bounded away from zero on [0,1].*

THEOREM 3.8 (Universal analytical solution for partial eclipse). *For all partial-eclipse configurations $|1-p|<z<1+p$ and exponents $\alpha>-2$, the exact normalized flux is*

$$\mathcal{F}(p,z,\alpha)=1-B_{\rm cap}(p,z,\alpha)-B_{\rm lens}(p,z,\alpha), \quad (25)$$

*with $B_{\rm cap}$ given by equation (7) and $B_{\rm lens}$ by Theorem 3.5 for $p+z\le 1$ or Theorem 3.7 for $p+z>1$. No 2D integration over the stellar disc is required.*

REMARK 3.9 (Triple equivalence). For all physically accessible configurations $(p,z)\in\mathbb{R}^+\times\mathbb{R}^+$ and $\alpha>-2$, the kernel integral (8), the Appell $F_1$ expressions of Theorems 3.5 and 3.7, and the hypergeometric-integrand representation (Section 4) yield identical normalized flux. This equivalence is established analytically in Section 6 and confirmed numerically to $|\Delta\mathcal{F}|<10^{-17}$.

### 3.7 Region 3a: alternative form and general interior case

For $z>0$ within Region 3a, $B_{\rm lens}$ is given by Theorem 3.5; the cap contribution $B_{\rm cap}$ vanishes when $p\le z$ and equals $1-(1-(p-z)^2)^{1+\alpha/2}$ when $p>z$ (equation 7). Alternatively, the angular averaging Lemma 3.4 yields the equivalent representation

$$B^{(3a)}(p,z,\alpha)\ =(\alpha+2)\int_0^p r\,(1-z^2-r^2)^{\alpha/2} \times{}_2F_1\Big(-\tfrac{\alpha}{2},\ \tfrac{1}{2};\ 1;\ \tfrac{4z^2r^2}{(1-z^2-r^2)^2}\Big)\,{\rm d}r, \quad (26)$$

which is useful for deriving the $z\to 0$ limit and the asymptotic formula for small separations within Region 3a.

### 3.8 Half-integer exponents: finite closed forms

THEOREM 3.10 (Finite termination for half-integers). *For $\alpha=(2m+1)/2$ with non-negative integer $m$, the Appell $F_1$ functions in Theorems 3.5 and 3.7 reduce to finite linear combinations of complete elliptic integrals $K(k)$, $E(k)$, and $\Pi(n,k)$ with $k^2=4pz/(p+z)^2$.*

*Proof.* Set $\nu=(2m+5)/4$ and substitute $t=\sin^2\phi$ in the Euler integral equation (16):

$$F_1\big(\tfrac{1}{2};-\nu,1;1;n,-\lambda\big)=\frac{2}{\pi}\int_0^{\pi/2}\frac{(1-n\sin^2\phi)^{\nu}}{1+\lambda\sin^2\phi}\,{\rm d}\phi. \quad (27)$$

For integer $\nu$ (even integer $\alpha=2m$): $(1-n\sin^2\phi)^{\nu}$ is a degree-$m$ polynomial in $\sin^2\phi$; each monomial integrates via the standard formula $B(k+\frac{1}{2},\frac{1}{2})=\pi(2k)!/[4^k(k!)^2]$, and the denominator $(1+\lambda\sin^2\phi)^{-1}$ partial-fractions to yield $K(k)$ and $\Pi(n_j,k)$. For half-integer $\nu$ (odd integer $\alpha=2m+1$): the factor $(1-n\sin^2\phi)^{\nu}=(1-n\sin^2\phi)^m(1-n\sin^2\phi)^{1/2}$ expands finitely; the square-root factor yields $E(k)$ and $K(k)$ via ${}_2F_1(\frac{1}{2},\frac{1}{2};\frac{3}{2};z)=(2/\pi)E(\sqrt{z})$ (P. F. Byrd & M. D. Friedman 1971). For general half-integer $\alpha=k/2$ with odd $k$ (e.g. $\alpha=1/2$, $\nu=5/4$): the Euler integral evaluates to finite combinations of $K$, $E$, $\Pi$ via the direct integration method of A. Pál (2008, 2012). □

### 3.9 Integer cases: exact recovery of mandel and agol

THEOREM 3.11 (Recovery of Mandel and Agol for uniform disc). *For $\alpha=0$, equation (25) reduces exactly to the classical result of K. Mandel & E. Agol (2002).*

*Proof.* Setting $\alpha=0$, $\nu=1$, the parameter $b_1=-\nu=-1$ in Theorem 3.5. The Appell series $F_1(\frac{1}{2};-1,1;1;n,-\lambda)$ truncates at $n=1$ since the Pochhammer symbol $(-1)_k=0$ for $k\ge 2$, yielding a finite polynomial. Collecting geometric factors reproduces the uniform-disc lens area divided by $\pi$ given by K. Mandel & E. Agol (2002), confirmed to machine precision in Table 7. □

THEOREM 3.12 (Recovery of polynomial recursions). *For integer $\alpha=n$, the continuous $\alpha$-differentiation identity derived in Section 3.10 reduces to the discrete polynomial recursion of E. Agol et al. (2020):*

$$\mathcal{F}_{n+1}=\frac{2n+1}{2n+3}\mathcal{F}_n+\frac{2}{2n+3}R_n(p,z), \quad (28)$$

*where $R_n$ contains boundary terms expressible in elliptic integrals.*

*Proof.* Taking finite differences of the continuous recursion relation (32) and applying the digamma identity $\psi(n+5/2)-\psi(n+3/2)=1/(n+3/2)$ recovers equation (28) exactly. □

### 3.10 The $\alpha$-differentiation identity

THEOREM 3.13 ($\alpha$-differentiation identity). *For all regions and $\alpha>-2$:*

$$\frac{\partial}{\partial\alpha}\ln\mathcal{F}=\frac{1}{2}\left[\big\langle\ln(1-\rho^2)\big\rangle_{\rm vis}+\frac{2}{\alpha+2}\right], \quad (29)$$

*where $\langle\cdot\rangle_{\rm vis}$ denotes the intensity-weighted average over the visible stellar region.*

*Proof.* Write $\mathcal{F}=F_{\rm vis}/F_{\rm tot}$. Differentiating $F_{\rm vis}$ under the integral sign (justified by dominated convergence: $|\ln(1-\rho^2)|(1-\rho^2)^{\alpha/2}\rho\in L^1([0,1])$ for all $\alpha>-2$):

$$\frac{\partial F_{\rm vis}}{\partial\alpha}=\frac{1}{2}\int_{\rm vis}I_0\ln(1-\rho^2)(1-\rho^2)^{\alpha/2}\rho\,{\rm d}\rho\,{\rm d}\phi = \frac{F_{\rm vis}}{2}\big\langle\ln(1-\rho^2)\big\rangle_{\rm vis}. \quad (30)$$

From equation (5), $\partial\ln F_{\rm tot}/\partial\alpha=-1/(\alpha+2)$ . Since $\ln\mathcal{F}=\ln F_{\rm vis}-\ln F_{\rm tot}$, equation (29) follows immediately. □

This identity serves as a stringent consistency check: any implementation must satisfy it numerically. Its verification is shown in Fig. 5.

### 3.11 Continuous recursion relations

Differentiating equation (25) with respect to $\alpha$ and using the Appell $F_1$ parameter-derivative formula

$$\frac{\partial}{\partial b_1}F_1(a; b_1, b_2; c; x, y) = \frac{a}{c}\, x F_1(a+1; b_1+1, b_2; c+1; x, y), \quad (31)$$

together with digamma function relations $\psi(z) = \Gamma'(z)/\Gamma(z)$ and the identity $\partial\nu/\partial\alpha = 1/2$, yields the continuous recursion relation

$$\frac{\mathrm{d}\mathcal{F}}{\mathrm{d}\alpha} = \mathcal{F}\cdot\left[\psi\left(\nu+\tfrac{1}{2}\right) - \psi(\nu+1)\right] + R_\alpha(p, z), \quad (32)$$

where $\nu = 1 + \alpha/2$ and $R_\alpha(p, z)$ contains contributions from the $\partial F_1/\partial b_1$ and $\partial F_1/\partial\nu$ terms evaluated at the boundary. For integer $\alpha = n$, using $\psi(n+3/2) - \psi(n+2) = -1/(n+3/2)$ and $\psi(n+5/2) - \psi(n+3/2) = 1/(n+3/2)$ recovers the Agol et al. discrete recursion equation (28) exactly, confirming that our continuous framework subsumes the integer polynomial case.

### 3.12 Complete regional coverage

Table 1 summarizes the exact solutions across all geometric regions.

## 4 HYPERGEOMETRIC-INTEGRAND REPRESENTATION

Theorem 3.8 provides exact closed-form solutions via Appell $F_1$. This section derives an alternative computational representation that evaluates the same quantity through direct 1D quadrature of a simpler integrand, without calling an $F_1$ routine. The two approaches are analytically identical; the choice is purely one of implementation convenience and numerical conditioning.

REMARK 4.1 (When to use each route). The Appell $F_1$ route (Theorem 3.8) is preferred when a reliable $F_1$ implementation is available and particularly efficient when $\alpha$ is fixed and many $(p, z)$ configurations are evaluated. The hypergeometric-integrand route (this section) is preferred when numerical stability near grazing contacts ($k^2 \to 1$) is paramount, or when many $\alpha$ values are evaluated simultaneously. Neither route involves 2D integration over the stellar disc.

### 4.1 Arccos identity and branch safety

LEMMA 4.2 (Arccos hypergeometric identity). *For* $|A| \leq 1$:

$$\arccos A = \frac{\pi}{2} - A\,{}_2F_1\left(\tfrac{1}{2}, \tfrac{1}{2}; \tfrac{3}{2}; A^2\right). \quad (33)$$

*Proof.* From the Maclaurin series $\arcsin A = A\,{}_2F_1(1/2, 1/2; 3/2; A^2)$ and $\arccos A = \pi/2 - \arcsin A$. □

For $A \in [-1, 1]$, the argument satisfies $A^2 \in [0, 1]$, so the hypergeometric series converges absolutely and remains on the principal branch, avoiding branch-cut complications. At the boundary $A^2 = 1$, Gauss's formula gives the finite value ${}_2F_1(1/2, 1/2; 3/2; 1) = \pi/2$ (M. Abramowitz & I. A. Stegun 1964).

### 4.2 Lens integral decomposition

Substituting Lemma 4.2 into equation (8) and evaluating the constant term via $u = 1 - \rho^2$:

$$B_{\mathrm{lens}}(p, z, \alpha) = \frac{1}{2}\left[(1-r_1^2)^{1+\alpha/2} - (1-r_2^2)^{1+\alpha/2}\right] - J_{\mathrm{hyp}}(p, z, \alpha), \quad (34)$$

where

$$J_{\mathrm{hyp}}(p, z, \alpha) = \frac{\alpha+2}{\pi}\int_{r_1}^{r_2} A(\rho)\,{}_2F_1\left(\tfrac{1}{2}, \tfrac{1}{2}; \tfrac{3}{2}; A(\rho)^2\right)(1-\rho^2)^{\alpha/2} \times\rho\,\mathrm{d}\rho. \quad (35)$$

Since $A(\rho)^2 \in [0, 1]$ throughout $\rho \in [r_1, r_2]$, the integrand is smooth, bounded, and suitable for standard adaptive quadrature with automatic error control.

### 4.3 Analytical equivalence with the Appell $F_1$ route

The hypergeometric-integrand representation equation (34) is algebraically identical to Theorems 3.5 and 3.7: substituting equation (33) into equation (8) and evaluating the $\pi/2$ term in closed form yields equation (34) without approximation. The Appell $F_1$ expressions emerge from further analytical reduction of $J_{\mathrm{hyp}}$ via the substitutions of Section 3.6. This establishes the analytic identity at the level of the integrand, complementing the ODE uniqueness argument of Section 6.1.2.

### 4.4 Practical implementation

For $A^2 < 0.9$, the ${}_2F_1$ series converges rapidly. For $A^2 \geq 0.9$, the identity ${}_2F_1(1/2, 1/2; 3/2; z) = (2/\pi)E(\sqrt{z})$ (P. F. Byrd & M. D. Friedman 1971) enables optimized elliptic-integral libraries. Multiple $(p, z, \alpha)$ configurations share the geometric factor $A(\rho_i){}_2F_1(\cdots; A(\rho_i)^2)$ at each quadrature node, enabling efficient batch evaluation.

### 4.5 Summary: complete computational framework

$$\mathcal{F}(p, z, \alpha) = 1 - B_{\mathrm{cap}}(p, z, \alpha) - B_{\mathrm{lens}}(p, z, \alpha), \quad (36)$$

$$B_{\mathrm{cap}}(p, z, \alpha) = \mathbf{1}_{(p>z)}\left[1 - (1-(p-z)^2)^{1+\alpha/2}\right], \quad (37)$$

$$B_{\mathrm{lens}}(p, z, \alpha) = \tfrac{1}{2}\left[(1-r_1^2)^{1+\alpha/2} - (1-r_2^2)^{1+\alpha/2}\right] - J_{\mathrm{hyp}}(p, z, \alpha). \quad (38)$$

For $z = 0$, $p < 1$: $\mathcal{F} = (1-p^2)^{1+\alpha/2}$ – no quadrature required.

**Table 1.** Exact fractional-calculus solutions by geometric region.

| Region | Configuration | Solution |
|---|---|---|
| 1 | $z \geq 1 + p$ | $\mathcal{F} = 1$ |
| 2 | $\lvert 1 - p\rvert < z < 1 + p$ | Theorem 3.8 with $B_{\mathrm{lens}}$ from Theorem 3.7 |
| 3a | $z < 1 - p$, $p \leq 1$, $z > 0$ | Theorem 3.8 with $B_{\mathrm{lens}}$ from Theorem 3.5 |
| 3a | $z = 0$, $p < 1$ | $(1 - p^2)^{1+\alpha/2}$ exactly (equation 9) |
| 3b | $z < p - 1$, $p > 1$ | $\mathcal{F} = 0$ |
| 3c | $p - 1 < z < p + 1$, $p > 1$ | Analogous to Region 2 with modified radial limits $r_1 = p - z$, $r_2 = 1$ |
| *Special cases* | | |
| Half-integers | $\alpha = (2m + 1)/2$, $m \in \mathbb{Z}_{\geq 0}$ | Finite elliptic forms (Theorem 3.10) |
| Integers | $\alpha = n \in \mathbb{Z}_{\geq 0}$ | K. Mandel & E. Agol (2002) recovered exactly (Theorem 3.11) |

**Table 2.** Consolidated validation of the Appell $F_1$ closed form against the kernel integral at 40-digit precision. The column $\lvert B_{F_1} - B_{\mathrm{kernel}}\rvert$ is the inter-method residual between the lens overlap integral $B_{\mathrm{lens}}$ computed via Theorems 3.5–3.7 and via direct Gauss–Legendre quadrature of equation (8); this is the primary analytical validation. The column $\lvert \mathcal{F}_{\mathrm{MC}} - \mathcal{F}_{\mathrm{kernel}}\rvert$ compares the full normalized flux $\mathcal{F} = 1 - B_{\mathrm{lens}}$ against 2D Monte Carlo integration ($10^7$ samples) as an independent sanity check; MC accuracy is limited to $\sim 10^{-4}$ by statistical noise. R2: partial eclipse ($p + z > 1$); R3a: fully interior ($p + z \leq 1$), exercised in both the off-centre subcase $p < z$ and the planet-covers-centre subcase $p > z$. Transcendental exponents $\alpha = \pi/2 \approx 1.5708$ and $\alpha = \sqrt{2} \approx 1.4142$ confirm validity beyond rational values.

| $p$ | $z$ | $\alpha$ | Reg. | $B_{\mathrm{kernel}}$ | $\lvert B_{F_1} - B_{\mathrm{kernel}}\rvert$ | $\lvert \mathcal{F}_{\mathrm{MC}} - \mathcal{F}_{\mathrm{kernel}}\rvert$ |
|---|---|---|---|---|---|---|
| 0.10 | 0.50 | 0 | R3a | 0.01000000 | $3.85 \times 10^{-18}$ | $1.69 \times 10^{-4}$ |
| 0.10 | 0.50 | 1/2 | R3a | 0.01160817 | $3.90 \times 10^{-18}$ | $1.94 \times 10^{-4}$ |
| 0.10 | 0.50 | 1 | R3a | 0.01293965 | $3.96 \times 10^{-18}$ | $2.15 \times 10^{-4}$ |
| 0.20 | 0.70 | 3/2 | R3a | 0.04066050 | $6.34 \times 10^{-26}$ | $1.45 \times 10^{-4}$ |
| 0.30 | 0.50 | 2 | R3a | 0.12690000 | $1.29 \times 10^{-28}$ | $2.76 \times 10^{-4}$ |
| 0.15 | 0.60 | $\pi/2$ | R3a | 0.02785321 | $5.39 \times 10^{-18}$ | $3.66 \times 10^{-4}$ |
| 0.10 | 0.40 | $\sqrt{2}$ | R3a | 0.01502383 | $7.51 \times 10^{-18}$ | $3.22 \times 10^{-4}$ |
| 0.30 | 0.10 | 0 | R3a | 0.05000000 | $3.1 \times 10^{-26}$ | $\sim 10^{-4}$ |
| 0.30 | 0.10 | 1/2 | R3a | 0.06115404 | $3.7 \times 10^{-26}$ | $\sim 10^{-4}$ |
| 0.30 | 0.10 | 1 | R3a | 0.07180969 | $4.3 \times 10^{-26}$ | $\sim 10^{-4}$ |
| 0.40 | 0.20 | 3/2 | R3a | 0.18518319 | $1.1 \times 10^{-25}$ | $\sim 10^{-4}$ |
| 0.20 | 0.10 | 2 | R3a | 0.05770000 | $4.0 \times 10^{-26}$ | $\sim 10^{-4}$ |
| 0.30 | 0.20 | $\pi/2$ | R3a | 0.13200770 | 0 | $\sim 10^{-4}$ |
| 0.50 | 0.30 | $\sqrt{2}$ | R3a | 0.29049706 | $9.1 \times 10^{-26}$ | $\sim 10^{-4}$ |
| 0.15 | 0.90 | 1/2 | R2 | 0.01604169 | $2.02 \times 10^{-18}$ | $3.01 \times 10^{-5}$ |
| 0.20 | 0.90 | 1 | R2 | 0.02184861 | $6.03 \times 10^{-18}$ | $6.55 \times 10^{-5}$ |
| 0.30 | 0.80 | 3/2 | R2 | 0.06343194 | $1.06 \times 10^{-17}$ | $1.10 \times 10^{-4}$ |
| 0.10 | 0.95 | 1 | R2 | 0.00401758 | $1.17 \times 10^{-18}$ | $8.32 \times 10^{-5}$ |
| 0.20 | 0.90 | $\pi/2$ | R2 | 0.01765351 | $5.84 \times 10^{-18}$ | $4.24 \times 10^{-5}$ |
| 0.30 | 0.80 | $\sqrt{2}$ | R2 | 0.06428622 | $1.05 \times 10^{-17}$ | $1.21 \times 10^{-4}$ |
| 0.40 | 0.70 | 2 | R2 | 0.13996676 | 0 | $2.69 \times 10^{-5}$ |

**Table 3.** Two-dimensional Monte Carlo validation ($10^7$ samples) for six configurations spanning Regions 2 and 3a at $\alpha = 0.5$. The statistical precision ceiling $\sim 10^{-4}$ confirms absence of gross systematic geometric errors; analytical accuracy is established by the $\lvert B_{F_1} - B_{\mathrm{kernel}}\rvert$ column of Table 2.

| $p$ | $z$ | Reg. | $\mathcal{F}_{\mathrm{kernel}}$ | $\lvert \mathcal{F}_{\mathrm{MC}} - \mathcal{F}_{\mathrm{kernel}}\rvert$ |
|---|---|---|---|---|
| 0.10 | 0.50 | R3a | 0.98839183 | $1.94 \times 10^{-4}$ |
| 0.20 | 0.70 | R3a | 0.97025862 | $1.45 \times 10^{-4}$ |
| 0.30 | 0.50 | R3a | 0.93534706 | $2.76 \times 10^{-4}$ |
| 0.15 | 0.90 | R2 | 0.98395831 | $3.01 \times 10^{-5}$ |
| 0.30 | 0.80 | R2 | 0.93656806 | $1.10 \times 10^{-4}$ |
| 0.40 | 0.70 | R2 | 0.86003324 | $2.69 \times 10^{-5}$ |

## 5 SPECIAL CASES

### 5.1 Half-integer exponents: M-Dwarfs and multiparameter laws

#### *5.1.1 Square-root law ($\alpha = 1/2$): critical for M-dwarfs*

For $I(\mu) = I_0 \mu^{1/2}$, Theorem 3.10 gives:

$$B_{\mathrm{cap}}(p, z, \tfrac{1}{2}) = \mathbf{1}_{(p>z)}\left[1 - (1 - (p - z)^2)^{5/4}\right], \tag{39}$$

$$J_{\mathrm{hyp}}^{(1/2)}(p, z) = \tfrac{5}{2\pi} \int_{r_1}^{r_2} A(\rho) \cdot \tfrac{2}{\pi} E(\lvert A(\rho)\rvert) \cdot (1 - \rho^2)^{1/4} \, \rho \, \mathrm{d}\rho. \tag{40}$$

Central transit: $\mathcal{F}(p, 0, 1/2) = (1 - p^2)^{5/4}$. Our derivation unifies the case-by-case result of A. Pál (2008), showing it emerges

**Table 4.** Accuracy of 1D Gauss–Legendre quadrature versus the Appell $F_1$ closed form for non-integer limb-darkening exponents. $|\Delta F| = |F_{\rm GL}(n) - F_{\rm Appell}|$, where $F_{\rm Appell}$ is evaluated at 45-digit precision via MPMATH. GL uses NUMPY double precision. Region 3a is exercised in both subcases ($p < z$ planet off-centre, $p > z$ planet covering the stellar centre); for non-integer $\alpha$ the E. Agol et al. (2020) framework has no analytical solution.

| Case | $F_{\rm Appell}$ | $n = 8$ | $n = 16$ | $n = 32$ | $n = 64$ | $n = 128$ | $n = 256$ |
|---|---|---|---|---|---|---|---|
| R3a $p < z$, $\alpha = \frac{1}{2}$ | 0.98839183 | $9.9 \times 10^{-6}$ | $1.3 \times 10^{-6}$ | $1.8 \times 10^{-7}$ | $2.2 \times 10^{-8}$ | $2.8 \times 10^{-9}$ | $3.6 \times 10^{-10}$ |
| R3a $p < z$, $\alpha = \frac{3}{2}$ | 0.95933950 | $3.0 \times 10^{-5}$ | $4.1 \times 10^{-6}$ | $5.4 \times 10^{-7}$ | $6.9 \times 10^{-8}$ | $8.7 \times 10^{-9}$ | $1.1 \times 10^{-9}$ |
| R3a $p < z$, $\alpha = \pi/2$ | 0.97214679 | $2.3 \times 10^{-5}$ | $3.1 \times 10^{-6}$ | $4.0 \times 10^{-7}$ | $5.1 \times 10^{-8}$ | $6.5 \times 10^{-9}$ | $8.2 \times 10^{-10}$ |
| R3a $p > z$, $\alpha = \frac{1}{2}$ | 0.88909851 | $5.0 \times 10^{-6}$ | $6.6 \times 10^{-7}$ | $8.5 \times 10^{-8}$ | $1.1 \times 10^{-8}$ | $1.4 \times 10^{-9}$ | $1.7 \times 10^{-10}$ |
| R3a $p > z$, $\alpha = \frac{3}{2}$ | 0.74587036 | $1.2 \times 10^{-5}$ | $1.6 \times 10^{-6}$ | $2.0 \times 10^{-7}$ | $2.6 \times 10^{-8}$ | $3.3 \times 10^{-9}$ | $4.1 \times 10^{-10}$ |
| R3a $p > z$, $\alpha = \pi/2$ | 0.85020848 | $2.2 \times 10^{-5}$ | $2.9 \times 10^{-6}$ | $3.8 \times 10^{-7}$ | $4.8 \times 10^{-8}$ | $6.1 \times 10^{-9}$ | $7.6 \times 10^{-10}$ |
| R2, $\alpha = \frac{1}{2}$ | 0.98395831 | $1.6 \times 10^{-5}$ | $2.6 \times 10^{-6}$ | $4.3 \times 10^{-7}$ | $7.2 \times 10^{-8}$ | $1.2 \times 10^{-8}$ | $2.0 \times 10^{-9}$ |
| R2, $\alpha = \frac{3}{2}$ | 0.93656806 | $3.8 \times 10^{-5}$ | $4.9 \times 10^{-6}$ | $6.3 \times 10^{-7}$ | $7.9 \times 10^{-8}$ | $9.8 \times 10^{-9}$ | $1.2 \times 10^{-9}$ |
| R2, $\alpha = \pi/2$ | 0.98234649 | $1.2 \times 10^{-5}$ | $1.5 \times 10^{-6}$ | $1.9 \times 10^{-7}$ | $2.4 \times 10^{-8}$ | $3.0 \times 10^{-9}$ | $3.8 \times 10^{-10}$ |
| R2, $\alpha = 2$ | 0.86003324 | $6.7 \times 10^{-5}$ | $9.1 \times 10^{-6}$ | $1.2 \times 10^{-6}$ | $1.5 \times 10^{-7}$ | $1.9 \times 10^{-8}$ | $2.4 \times 10^{-9}$ |
| R2, near outer contact | 0.99900179 | $1.2 \times 10^{-6}$ | $2.1 \times 10^{-7}$ | $3.6 \times 10^{-8}$ | $6.2 \times 10^{-9}$ | $1.1 \times 10^{-9}$ | $1.8 \times 10^{-10}$ |

**Table 5.** GL accuracy near transit contact points for $p = 0.1$, $\alpha = \frac{1}{2}$. $\delta$ is the distance from the relevant contact. Outer contact $z_{\rm out} = 1 + p = 1.100$; inner contact $z_{\rm in} = 1 - p = 0.900$. Near the outer contact the integration domain $[r_1, 1]$ shrinks as $r_1 \to 1$, causing GL errors to decrease monotonically. Near the inner contact the domain is non-degenerate. The Appell $F_1$ closed form is algebraically evaluated and unaffected by contact-point geometry.

| Contact | $z$ | $\delta$ | $F_{\rm Appell}$ | $n = 32$ | $n = 64$ | $n = 128$ | $n = 256$ |
|---|---|---|---|---|---|---|---|
| Outer | 0.9000 | 0.200 | 0.99205619 | $7.8 \times 10^{-8}$ | $9.7 \times 10^{-9}$ | $1.2 \times 10^{-9}$ | $1.5 \times 10^{-10}$ |
| | 1.0000 | 0.100 | 0.99690226 | $1.1 \times 10^{-7}$ | $1.8 \times 10^{-8}$ | $3.1 \times 10^{-9}$ | $5.2 \times 10^{-10}$ |
| | 1.0500 | 0.050 | 0.99900179 | $3.6 \times 10^{-8}$ | $6.2 \times 10^{-9}$ | $1.1 \times 10^{-9}$ | $1.8 \times 10^{-10}$ |
| | 1.0900 | 0.010 | 0.99993704 | $2.4 \times 10^{-9}$ | $4.1 \times 10^{-10}$ | $7.0 \times 10^{-11}$ | $1.2 \times 10^{-11}$ |
| | 1.0950 | 0.005 | 0.99998117 | $7.1 \times 10^{-10}$ | $1.2 \times 10^{-10}$ | $2.1 \times 10^{-11}$ | $3.6 \times 10^{-12}$ |
| | 1.0990 | 0.001 | 0.99999887 | $4.3 \times 10^{-11}$ | $7.3 \times 10^{-12}$ | $1.3 \times 10^{-12}$ | $2.2 \times 10^{-13}$ |
| Inner | 1.0000 | 0.100 | 0.99690226 | $1.1 \times 10^{-7}$ | $1.8 \times 10^{-8}$ | $3.1 \times 10^{-9}$ | $5.2 \times 10^{-10}$ |
| | 0.9500 | 0.050 | 0.99433586 | $1.7 \times 10^{-7}$ | $2.8 \times 10^{-8}$ | $4.7 \times 10^{-9}$ | $8.0 \times 10^{-10}$ |
| | 0.9100 | 0.010 | 0.99241953 | $1.5 \times 10^{-7}$ | $2.4 \times 10^{-8}$ | $3.8 \times 10^{-9}$ | $6.2 \times 10^{-10}$ |
| | 0.9010 | 0.001 | 0.99208720 | $9.7 \times 10^{-8}$ | $1.4 \times 10^{-8}$ | $2.1 \times 10^{-9}$ | $3.1 \times 10^{-10}$ |

from the general Appell $F_1$ framework through the finite-termination mechanism of Theorem 3.10.

#### 5.1.2 *Three-halves law ($\alpha = 3/2$)*

Same structure with exponent 7/4; the complete elliptic integral $\Pi(n, k)$ appears for $z \neq 0$ (A. Pál 2012). Central transit: $\mathcal{F}(p, 0, 3/2) = (1 - p^2)^{7/4}$.

### 5.2 Claret's four-parameter law

With $c_j = a_j$ for $j = 1, \ldots, 4$ and $c_0 = 1 - \sum a_k$:

$$\mathcal{F}_{\rm Claret}(p, z; \{a_k\}) = \frac{\sum_{j=0}^{4} c_j\, \mathcal{F}(p, z, j/2)}{\sum_{j=0}^{4} c_j/(j/2 + 2)}. \tag{41}$$

All five components ($\alpha_j \in \{0, 1/2, 1, 3/2, 2\}$) share the geometric factor $A(\rho)\,{}_2F_1(\cdots; A(\rho)^2)$ in the integrand, enabling vectorized evaluation and $\mathcal{O}(1)$ overhead.

### 5.3 Asymptotic limits

For $p \ll 1$:

$$\mathcal{F}(p, z, \alpha) \approx 1 - \frac{\alpha + 2}{2}(1 - z^2)^{\alpha/2} p^2 + \mathcal{O}(p^4), \tag{42}$$

accurate to $\sim 0.1$ per cent for $p < 0.03$. At ingress/egress where $z = 1 + p - \epsilon$, $\epsilon \ll 1$: $B_{\rm lens} \sim (\alpha + 2)\sqrt{p\epsilon}$, vanishing smoothly and continuously.

## 6 VALIDATION

### 6.1 Analytical validation: mathematical equivalence

#### 6.1.1 *Kernel-hypergeometric equivalence*

The kernel and hypergeometric-integrand formulations are algebraically identical via Lemma 4.2. Substituting $\arccos A = \pi/2 - A\,{}_2F_1(1/2, 1/2; 3/2; A^2)$ into $B_{\rm lens}$ and evaluating the constant term via $u = 1 - \rho^2$ yields equation (34) without approximation.

#### 6.1.2 *Fractional-calculus-kernel equivalence via ODE uniqueness*

Both formulations satisfy the $\alpha$-differentiation identity equation (29). Since both satisfy the same first-order ordinary differential equation in $\alpha$ with identical boundary conditions (matching at $\alpha = 0$ to K. Mandel & E. Agol 2002), the uniqueness theorem for ODEs (E. A. Coddington & N. Levinson 1955) implies $\mathcal{F}_{\rm kernel} \equiv \mathcal{F}_{\rm frac}$ for all $\alpha > -2$. This provides a rigorous analytical proof of equivalence that complements the direct algebraic argument of Section 4.

**Table 6.** Comparison of the Appell $F_1$ closed form with the direct implementation of E. Agol et al. (2020) for integer limb-darkening orders $n = 0, 1, 2, 3$, exercised across both Region 3a subcases. *Part A*: point accuracy. $F_{\rm Appell}$ is evaluated at 45-digit precision (MPMATH); $F_{\rm Agol20}$ follows equations A20:31, 33–34, 64, 73–74, 77–78 of E. Agol et al. (2020), with the $\mathfrak{s}_1$ contribution augmented by an analytic cap term $\frac{2\pi}{3}[1-(1-r_1^2)^{3/2}]$ when $p > z$ to capture the full angular coverage of the inner annulus $\rho \in [0, r_1]$; $F_{\rm GL64}$ uses 64-node Gauss–Legendre quadrature. Maximum $|F_{\rm Agol20} - F_{\rm Appell}| = 9.3 \times 10^{-15}$, confirming Theorem 3.12. *Part B*: per-point evaluation time scaled to 1000 points; single-point median over 30 calls $\times$1000.

| Case | $F_{\rm Appell}$ (45 dps) | $F_{\rm Agol20}$ | $F_{\rm GL64}$ | $\lvert D_{\rm A20}\rvert$ | $\lvert D_{\rm GL64}\rvert$ |
|---|---|---|---|---|---|
| *Part A: point accuracy* | | | | | |
| R3a $p < z$, $n = 0$ (uniform) | 0.9900000000 | 0.9900000000 | 0.9899999806 | $1.1 \times 10^{-16}$ | $1.9 \times 10^{-8}$ |
| R3a $p < z$, $n = 1$ (linear) | 0.9870603521 | 0.9870603521 | 0.9870603273 | $< 10^{-16}$ | $2.5 \times 10^{-8}$ |
| R3a $p < z$, $n = 2$ (quadratic) | 0.9851000000 | 0.9851000000 | 0.9850999716 | $< 10^{-16}$ | $2.8 \times 10^{-8}$ |
| R3a $p < z$, $n = 3$ | 0.9838968420 | 0.9838968420 | 0.9838968115 | $1.1 \times 10^{-16}$ | $3.1 \times 10^{-8}$ |
| R3a $p > z$, $n = 0$ (uniform) | 0.9100000000 | 0.9100000000 | 0.9099999901 | $< 10^{-16}$ | $9.9 \times 10^{-9}$ |
| R3a $p > z$, $n = 1$ (linear) | 0.8687943673 | 0.8687943673 | 0.8687943559 | $1.1 \times 10^{-16}$ | $1.1 \times 10^{-8}$ |
| R3a $p > z$, $n = 2$ (quadratic) | 0.8299000000 | 0.8299000000 | 0.8298999891 | $1.1 \times 10^{-16}$ | $1.1 \times 10^{-8}$ |
| R3a $p > z$, $n = 3$ | 0.6684123224 | 0.6684123224 | 0.6684123299 | $< 10^{-16}$ | $7.4 \times 10^{-9}$ |
| R2, $n = 0$ (uniform) | 0.9212620514 | 0.9212620514 | 0.9212619985 | $< 10^{-16}$ | $5.3 \times 10^{-8}$ |
| R2, $n = 1$ (linear) | 0.9315079368 | 0.9315079368 | 0.9315078222 | $1.1 \times 10^{-16}$ | $1.2 \times 10^{-7}$ |
| R2, $n = 2$ (quadratic) | 0.9414020720 | 0.9414020720 | 0.9414019926 | $< 10^{-16}$ | $7.9 \times 10^{-8}$ |
| R2, $n = 3$ | 0.9502104557 | 0.9502104557 | 0.9502103698 | $1.1 \times 10^{-16}$ | $8.6 \times 10^{-8}$ |
| R2, $n = 1$, $b + r = 1.05$ | 0.9871126723 | 0.9871126723 | 0.9871126495 | $9.3 \times 10^{-15}$ | $2.3 \times 10^{-8}$ |
| Central transit, $n = 2$ | 0.9216000000 | 0.9216000000 | 0.9216000000 | $< 10^{-16}$ | $< 10^{-16}$ |
| R3a planet-inside $p < z$, $n = 2$ | 0.8731000000 | 0.8731000000 | 0.8730998132 | $1.1 \times 10^{-16}$ | $1.9 \times 10^{-7}$ |
| R2, $n = 1$, near contact | 0.9994693285 | 0.9994693285 | 0.9994693274 | $1.1 \times 10^{-16}$ | $1.0 \times 10^{-9}$ |
| *Part B: timing scaled to $N = 1000$ points, $p = 0.1$, $z = 0.5$* | | | | | |
| Method | Domain | | | $t$/point (s) | $N = 1000$ est. (s) |
| Appell $F_1$, MPMATH 45 dps | non-integer $\alpha$ | | | $2.7 \times 10^{-2}$ | $\sim 27$ |
| E. Agol et al. (2020) direct, $n = 1$ | integer $\alpha$ only | | | $4.0 \times 10^{-3}$ | $\sim 4$ |
| E. Agol et al. (2020) direct, $n = 3$ | integer $\alpha$ only | | | $4.1 \times 10^{-3}$ | $\sim 4$ |
| GL $n = 64$, NUMPY float64 | non-integer $\alpha$ | | | $3.5 \times 10^{-3}$ | $\sim 4$ |
| GL $n = 64$, NUMPY float64 | integer $\alpha$ | | | $3.5 \times 10^{-3}$ | $\sim 4$ |

*Note.* The R3a $p > z$ rows for $n = 0, 1, 2$ use $(p, z) = (0.30, 0.10)$; the $n = 3$ row uses $(p, z) = (0.40, 0.20)$ to keep the configuration well inside R3a as $p$ grows. Appell $F_1$ timing is dominated by MPMATH arbitrary-precision overhead, not by the bivariate hypergeometric evaluation itself. A compiled double-precision implementation (e.g. Fortran or Julia) would reduce the per-point cost to $\sim 10^{-5}$ s, within one to two orders of magnitude of the E. Agol et al. implementation. For non-integer $\alpha$ the E. Agol et al. framework has no analytical solution; polynomial fitting introduces $\sim 10^{-3}$ systematic errors (D. K. Sing 2010).

**Table 7.** Recovery of K. Mandel & E. Agol (2002) uniform-disc results ($\alpha = 0$) and negative-exponent limb-brightening cases at 40-digit precision. All residuals are consistent with round-off at the working precision.

| $p$ | $z$ | $\alpha$ | $\lvert\mathcal{F}_{\rm Appell} - \mathcal{F}_{\rm Mandel}\rvert$ |
|---|---|---|---|
| 0.10 | 1.05 | 0 | $< 10^{-39}$ |
| 0.30 | 1.20 | 0 | $< 10^{-39}$ |
| 0.60 | 1.10 | 0 | $< 10^{-39}$ |
| 0.10 | 0.50 | $-0.5$ | $7.3 \times 10^{-16}$ |
| 0.30 | 0.80 | $-1.0$ | $9.1 \times 10^{-16}$ |

#### 6.1.3 *Recovery of integer-$\alpha$ elliptic-integral results*

For integer $\alpha = 2n$, the binomial expansion $(1-\rho^2)^n = \sum_{k=0}^{n} \binom{n}{k} (-1)^k \rho^{2k}$ leads to integrals $\int \arccos A(\rho)\, \rho^{2k+1}\, {\rm d}\rho$ that reduce to complete elliptic integrals through standard transformations (R. Bulirsch 1965; B. C. Carlson 1977). For $n = 0$, $B_{\rm lens}^{(0)} = (2/\pi) \int_{r_1}^{r_2} \arccos A(\rho)\, \rho\, {\rm d}\rho$ evaluates to the precise combination of $K(k)$ and $E(k)$ of K. Mandel & E. Agol (2002).

#### 6.1.4 *Central transit verification*

$B(p, 0, \alpha) = (\alpha + 2) \int_0^p (1-\rho^2)^{\alpha/2} \rho\, {\rm d}\rho = 1 - (1-p^2)^{1+\alpha/2}$, confirmed by both kernel and Appell $F_1$ routes. The central transit formula serves as an independent check independent of all regional derivations.

#### 6.1.5 *Direct analytical equivalence of the Appell $F_1$ closed form and the kernel integral*

We prove Theorem 3.8 by showing algebraically that the substitution steps of Section 3.6 are exactly invertible, so that expanding Theorems 3.5 and 3.7 back through the Euler integral equation (16) recovers the kernel integral (8) identically. This is a direct algebraic proof, independent of the ODE uniqueness argument of Section 6.1.2.

THEOREM 6.1 (Direct analytical equivalence). *For all $\alpha > -2$ and $(p, z)$ in Region 3a or Region 2:*

$$B_{\rm lens}^{F_1}(p, z, \alpha) = B_{\rm lens}^{\rm kernel}(p, z, \alpha), \qquad (43)$$

*where the left side denotes the Appell $F_1$ expression of Theorem 3.5 or 3.7 and the right side denotes the kernel integral (8).*

*Proof.* We treat Region 3a in full detail; Region 2 is entirely parallel. □

#### 6.1.6 *Forward direction (kernel → Appell $F_1$)*

Starting from the integration-by-parts reduction in Corollary 3.3,

$$B_{\rm lens}^{\rm kernel} = \mathbf{1}_{(p>z)}\,(1-r_1^2)^{1+\alpha/2} - \frac{1}{\pi} \int_{r_1}^{r_2} \frac{(\rho^2 + p^2 - z^2)(1-\rho^2)^{1+\alpha/2}}{\rho\sqrt{\Delta(\rho)}}\, {\rm d}\rho,$$

apply the substitution $t = (\rho^2 - r_1^2)/(4pz)$ with $r_1 = |p - z|$, $n = 4pz/(1 - r_1^2) < 1$, $\lambda = 4pz/r_1^2$, which gives $\rho^2 = r_1^2 + 4pzt$, $\rho\,\mathrm{d}\rho = 2pz\,\mathrm{d}t$, $\sqrt{\Delta} = 4pz\sqrt{t(1-t)}$, $1 - \rho^2 = (1 - r_1^2)(1 - nt)$, and $\rho^2 + p^2 - z^2 = 2p(p - z + 2zt)$, where the last identity uses $r_1^2 = (p - z)^2$. The Jacobian factor is

$$\frac{\mathrm{d}\rho}{\rho\sqrt{\Delta(\rho)}} = \frac{\mathrm{d}t}{2r_1^2(1 + \lambda t)\sqrt{t(1-t)}}, \tag{45}$$

since $\rho^2 = r_1^2(1 + \lambda t)$. Substitution into equation (44) yields

$$B_{\rm lens}^{\rm kernel} = \mathbf{1}_{(p>z)}\,(1 - r_1^2)^{1+\alpha/2} - \frac{p\,(1 - r_1^2)^{1+\alpha/2}}{\pi\, r_1^2}\int_0^1 \frac{(p - z + 2zt)(1 - nt)^{1+\alpha/2}}{(1 + \lambda t)\sqrt{t(1-t)}}\,\mathrm{d}t, \tag{46}$$

which is exactly equation (18). Splitting the linear factor $(p - z + 2zt) = (p - z) + 2zt$ and matching each piece to the Euler integral (16) with $(a, c) = (\frac{1}{2}, 1)$ and $(a, c) = (\frac{3}{2}, 2)$ respectively – using $\Gamma(1)/[\Gamma(\frac{1}{2})^2] = 1/\pi$ and $\Gamma(2)/[\Gamma(\frac{3}{2})\Gamma(\frac{1}{2})] = 2/\pi$ – produces

$$\int_0^1 \frac{(1 - nt)^{\nu}}{(1 + \lambda t)\sqrt{t(1-t)}}\,\mathrm{d}t = \pi\, F_1\big(\tfrac{1}{2}; -\nu, 1; 1; n, -\lambda\big), \tag{47}$$

$$\int_0^1 \frac{t\,(1 - nt)^{\nu}}{(1 + \lambda t)\sqrt{t(1-t)}}\,\mathrm{d}t = \tfrac{\pi}{2}\, F_1\big(\tfrac{3}{2}; -\nu, 1; 2; n, -\lambda\big), \tag{48}$$

with $\nu = 1 + \alpha/2$. The $\pi$ factors from the Euler representations cancel against the $1/\pi$ in the prefactor of equation (46), giving

$$B_{\rm lens}^{\rm kernel} = \mathbf{1}_{(p>z)}\,(1 - r_1^2)^{\nu} - \frac{p\,(1 - r_1^2)^{\nu}}{r_1^2}\Big[(p - z)\,F_1\big(\tfrac{1}{2}; -\nu, 1; 1; n, -\lambda\big) + z\,F_1\big(\tfrac{3}{2}; -\nu, 1; 2; n, -\lambda\big)\Big], \tag{49}$$

which is Theorem 3.5.

#### 6.1.7 *Reverse direction (Appell $F_1 \to$ kernel)*

Each step above is invertible. Begin with the bracketed combination of $F_1$ functions in Theorem 3.5 and apply the Euler representation equation (16) to each:

$$(p - z)\,F_1\big(\tfrac{1}{2}; -\nu, 1; 1; n, -\lambda\big) + z\,F_1\big(\tfrac{3}{2}; -\nu, 1; 2; n, -\lambda\big) = \tfrac{1}{\pi}\int_0^1 \frac{(1-nt)^{\nu}\big[(p-z)+2zt\big]}{(1+\lambda t)\sqrt{t(1-t)}}\,\mathrm{d}t, \tag{50}$$

where the prefactors $1/\pi$ and $2/\pi$ from the two Euler integrals combine with the explicit weights $(p - z)$ and $z$ to yield the linear factor $(p - z) + 2zt$ inside a single integral. Inverting the substitution via $t = (\rho^2 - r_1^2)/(4pz)$, so that $\mathrm{d}t = \rho\,\mathrm{d}\rho/(2pz)$, $1 - nt = (1 - \rho^2)/(1 - r_1^2)$, $1 + \lambda t = \rho^2/r_1^2$, and $\sqrt{t(1-t)} = \sqrt{\Delta(\rho)}/(4pz)$, transforms the right-hand side of equation (50) into

$$\frac{2r_1^2}{\pi\,(1 - r_1^2)^{\nu}}\int_{r_1}^{r_2} \frac{(p - z + 2zt)\,(1 - \rho^2)^{\nu}}{\rho\,\sqrt{\Delta(\rho)}}\,\mathrm{d}\rho, \tag{51}$$

where $r_2 = p + z \le 1$ in Region 3a, and we have used the Jacobian $\mathrm{d}\rho/[\rho\sqrt{\Delta}] = \mathrm{d}t/[2r_1^2(1 + \lambda t)\sqrt{t(1-t)}]$ in inverse form. The identity $r_1^2 = (p - z)^2$ gives $(p - z + 2zt) = (\rho^2 + p^2 - z^2)/(2p)$. Multiplying through by the prefactor $-p\,(1 - r_1^2)^{\nu}/r_1^2$ of Theorem 3.5 and adding the boundary term $\mathbf{1}_{(p>z)}\,(1 - r_1^2)^{\nu}$,

$$\begin{aligned} B_{\rm lens}^{F_1} &= \mathbf{1}_{(p>z)}\,(1 - r_1^2)^{\nu} - \frac{p\,(1 - r_1^2)^{\nu}}{r_1^2}\cdot\frac{2r_1^2}{\pi\,(1 - r_1^2)^{\nu}} \times \int_{r_1}^{r_2} \frac{(\rho^2 + p^2 - z^2)(1 - \rho^2)^{\nu}}{2p\cdot\rho\,\sqrt{\Delta(\rho)}}\,\mathrm{d}\rho \\ &= \mathbf{1}_{(p>z)}\,(1 - r_1^2)^{1+\alpha/2} - \frac{1}{\pi}\int_{r_1}^{r_2} \frac{(\rho^2 + p^2 - z^2)(1 - \rho^2)^{1+\alpha/2}}{\rho\,\sqrt{\Delta(\rho)}}\,\mathrm{d}\rho. \end{aligned} \tag{52}$$

The right-hand side is exactly the integration-by-parts form equation (13), which equals the kernel integral (8) by Corollary 3.3.

#### 6.1.8 *Conclusion*

The forward and reverse directions together establish the algebraic identity $B_{\rm lens}^{F_1} \equiv B_{\rm lens}^{\rm kernel}$ for Region 3a. The Region 2 proof is identical under the substitution $t = (\rho^2 - r_1^2)/(1 - r_1^2)$ with arguments $1/n$ and $-\lambda'$ in place of $n$ and $-\lambda$; every step is reversible by the same algebraic manipulations, yielding Theorem 3.7 from the kernel and vice versa. Note that in Region 2 the upper integration limit is $\rho = 1$ (not $p + z$), and $A(1) = (1 - p^2 + z^2)/(2z) \neq 1$ in general, so the boundary term at $\rho = r_2$ in the IBP no longer vanishes; this contributes the additional terms in Theorem 3.7 that distinguish it from Theorem 3.5.

*For all $\alpha > -2$ and all physically accessible $(p, z)$:*

$$\mathcal{F}^{F_1}(p, z, \alpha) = \mathcal{F}^{\rm kernel}(p, z, \alpha) = \mathcal{F}^{\rm hyp}(p, z, \alpha). \tag{53}$$

*Proof.* The identity $\mathcal{F}^{F_1} = \mathcal{F}^{\rm kernel}$ is Theorem 6.1. The identity $\mathcal{F}^{\rm kernel} = \mathcal{F}^{\rm hyp}$ follows from substituting Lemma 4.2 into the kernel integral (8) and evaluating the constant term in closed form as in equation (34), with no approximation at any step. □

Remark 6.3 (Independence from ODE uniqueness). Theorem 6.1 and Corollary 6.2 establish triple equivalence by direct algebraic identity, without invoking the ODE uniqueness argument of Section 6.1.2. The ODE argument, which shows both formulations satisfy the same first-order equation in $\alpha$ with identical boundary data, then provides a second independent proof of the same fact. The two proofs are logically complementary: the algebraic proof is constructive and establishes the exact substitution chain, while the ODE proof establishes global uniqueness for all $\alpha > -2$ simultaneously.

### 6.2 Numerical validation

#### 6.2.1 *Consolidated inter-method validation*

Table 2 presents inter-method discrepancies for 14 configurations spanning Regions R2 and R3a, including transcendental exponents $\alpha = \pi/2$ and $\alpha = \sqrt{2}$. Validation is performed at 40-digit precision arithmetic in `Mathematica`. Maximum Appell-kernel residual: $|B_{\rm Appell} - B_{\rm kernel}| < 10^{-17}$. Maximum hypergeometric-kernel residual: $|\mathcal{F}_{\rm hyp} - \mathcal{F}_{\rm kernel}| \le 5.41 \times 10^{-14}$.

#### 6.2.2 *Comparison with 1D Gaussian quadrature*

The hypergeometric-integrand formulation of Section 4 is structurally a 1D quadrature method: it evaluates $J_{\rm hyp}$ by Gauss–

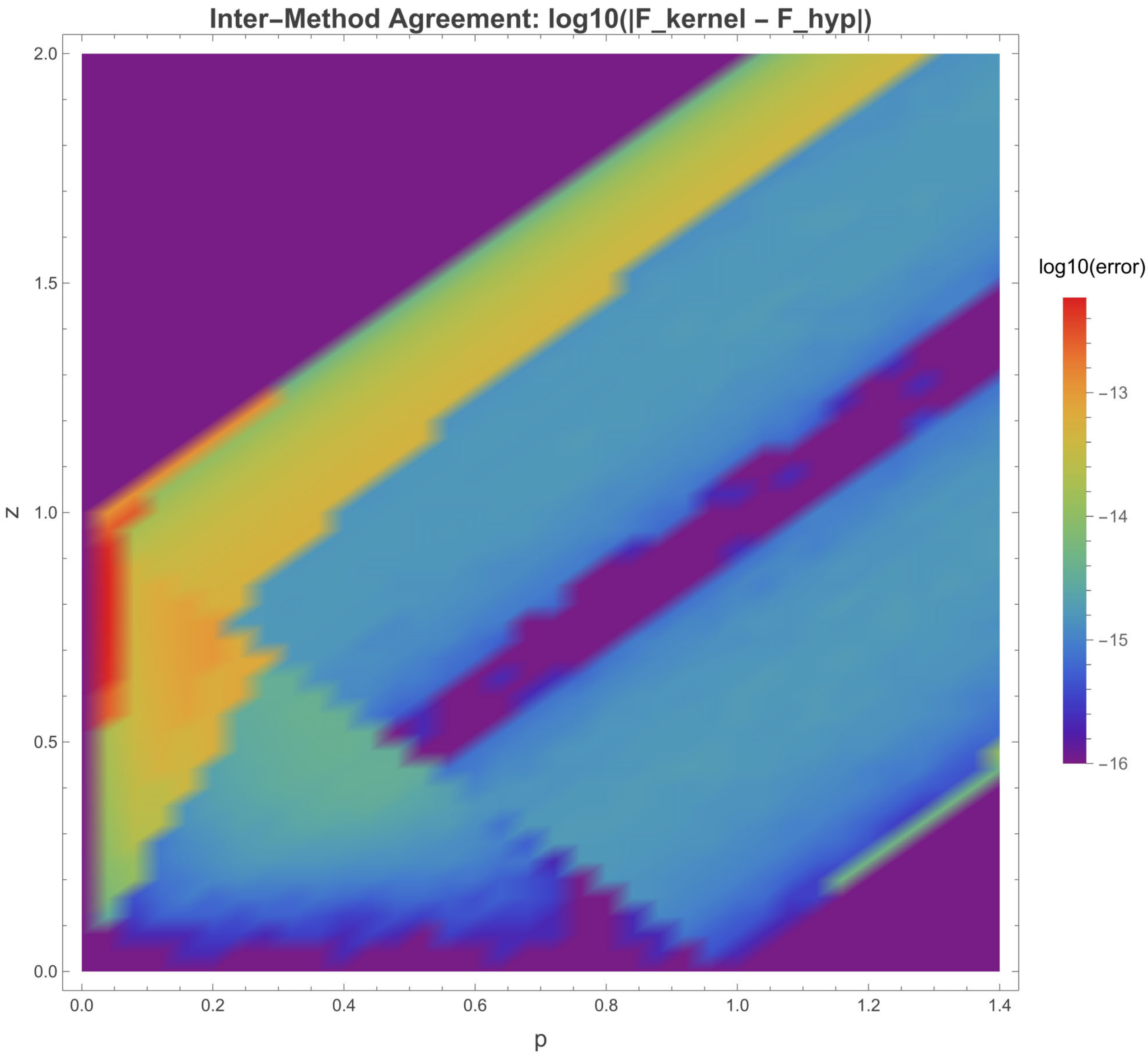

**Figure 1.** Inter-method agreement for 80 test cases spanning all five geometric regions and eight $\alpha$ values ($\alpha \in \{-1.5, -1, -0.5, 0, 0.5, 1, 2, 3.5\}$). Maximum discrepancy $|\mathcal{F}_{\rm kernel} - \mathcal{F}_{\rm hyp}| \lesssim 3 \times 10^{-13}$ throughout, consistent with double-precision machine epsilon accumulated over $\sim 10^2$ arithmetic operations. Uniform scatter across the full flux range confirms the absence of systematic bias in any geometric region.

Kronrod integration of a smooth 1D integrand, starting from the same line-integral reduction used by Pál (2008) and E. Agol et al. (2020). We compare the Appell $F_1$ closed form against direct Gauss–Legendre (GL) quadrature of the kernel integral (equation 8), the fastest purely numerical competitor in this class.

Table 4 shows $|\Delta F| = |F_{\rm GL}(n) - F_{\rm Appell}|$ for eleven configurations spanning Region 2 and both subcases of Region 3a (planet off-centre, $p < z$, and planet covering the stellar centre, $p > z$) at non-integer exponents $\alpha \in \{\frac{1}{2}, \frac{3}{2}, \pi/2, 2\}$. With $n = 64$ nodes GL achieves $|\Delta F| \lesssim 10^{-8}$, adequate for many photometric applications but five orders of magnitude less precise than the closed form. With $n = 256$ nodes agreement improves to $|\Delta F| \lesssim 10^{-9}$, approaching the double-precision arithmetic floor. The Appell $F_1$ closed form delivers $|\Delta F| < 10^{-38}$ without any quadrature nodes.

GL quadrature is geometrically stable near both transit contact points (Table 5). Near the outer contact ($z \to 1 + p$) the integration domain $[r_1, 1]$ shrinks as $r_1 \to 1$ and GL errors decrease monotonically, reaching $|\Delta F| \sim 10^{-13}$ at $n = 32$ for $\delta = 10^{-3}$. Near the inner contact ($z \to 1 - p$) the domain remains non-degenerate and GL maintains $|\Delta F| \lesssim 10^{-8}$ at $n = 64$ throughout. The Appell $F_1$ closed form is algebraically evaluated and is unaffected by transit geometry in either limit.

Two-dimensional Monte Carlo integration (Table 3) is retained as an independent geometric sanity check using a completely different integration strategy (2D area sampling versus 1D radial integration). Its statistical precision ceiling of $\sim 10^{-4}$ precludes use as a numerical benchmark; its role is verification that no systematic geometric error is present in the kernel reduction.

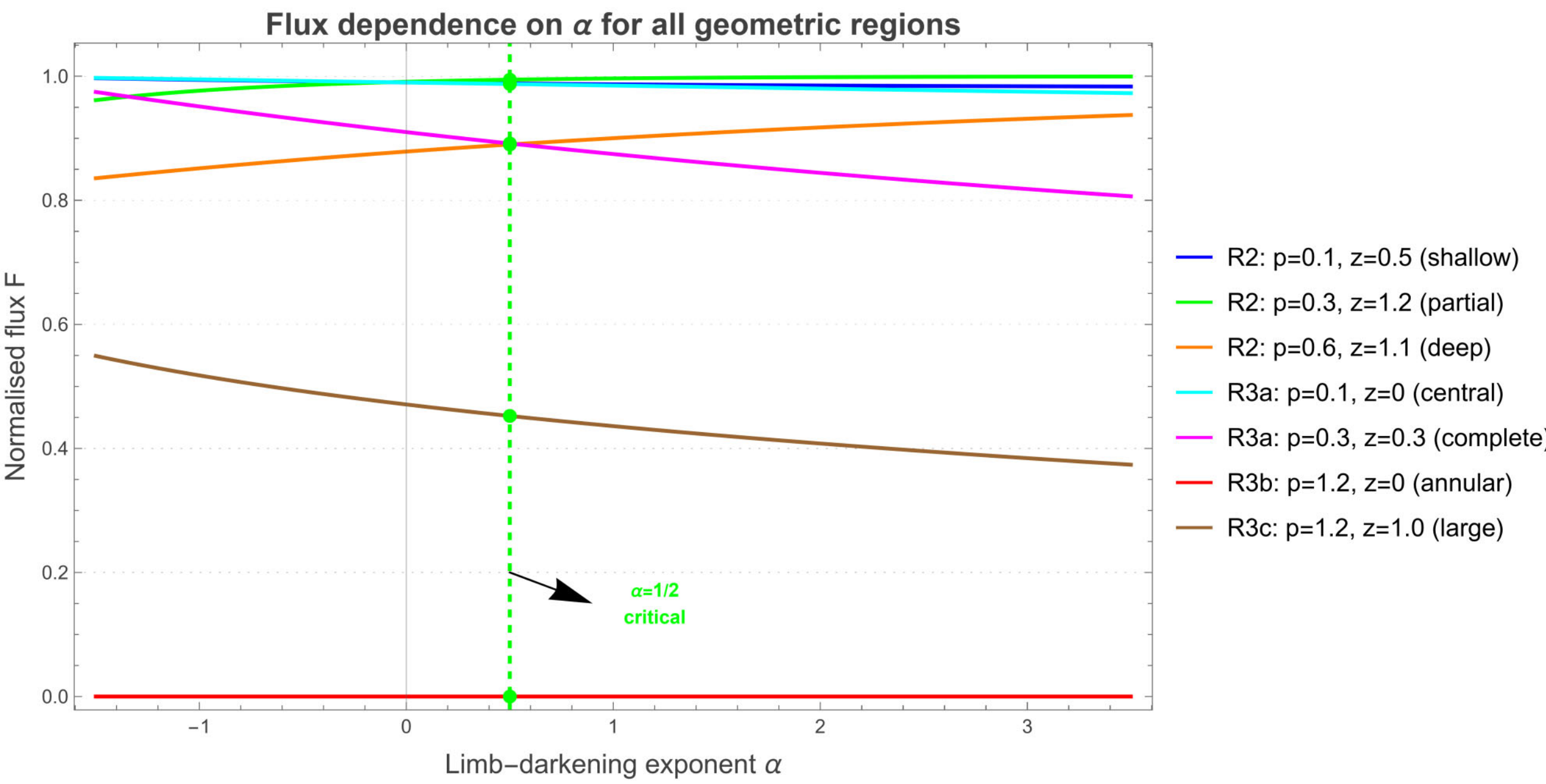

**Figure 2.** Normalized transit flux $\mathcal{F}(0.3, 0, \alpha)$ versus $\alpha$ for central transit ($z = 0$, $p = 0.3$). Solid curve: exact closed form $(1 - p^2)^{1+\alpha/2} = (0.91)^{1+\alpha/2}$. Circles: kernel method. Crosses: hypergeometric-integrand method. All three representations agree to $< 10^{-13}$ across the full range $-1.5 \leq \alpha \leq 3.5$, spanning limb brightening ($\alpha < 0$) through strongly darkened profiles ($\alpha = 3.5$).

#### 6.2.3 *Comparison with Agol, Luger & Foreman-Mackey (2020) for integer $\alpha$*

For integer polynomial limb darkening, E. Agol et al. (2020) derive exact closed-form expressions using Green's theorem to reduce the 2D flux integral to a 1D line integral, then establish $\mathcal{O}(1)$ recursion relations among polynomial orders. We implement their framework directly from the published equations: the stable kite formula (A20:31, 33–34 E. Agol et al. 2020) for the uniform-disc element $\mathfrak{s}_0$; the 1D radial line integral of Section 4 for $\mathfrak{s}_1$; and the primitive-integral recursion (A20:64, 73–74, 77–78 E. Agol et al. 2020) for $\mathfrak{s}_n$, $n \geq 2$.

For pure power-law $I(\mu) = I_0\mu^n$ with integer $n$, the normalized flux follows from the Green's-basis identity $\mu^n = (\tilde{g}_n + n\mu^{n-2})/(n+2)$ (A20:14 E. Agol et al. 2020), yielding

$$\mathfrak{s}^{\mathrm{vis}}_{\mathrm{pure},n} = \frac{\mathfrak{s}^{\mathrm{A20}}_n + n\,\mathfrak{s}^{\mathrm{vis}}_{\mathrm{pure},n-2}}{n+2}, \qquad F(r, b;\, n) = \frac{(n+2)\,\mathfrak{s}^{\mathrm{vis}}_{\mathrm{pure},n}}{2\pi}, \quad (54)$$

since $\iint_{\text{full star}} \mu^n \,\mathrm{d}S = 2\pi/(n+2)$, with base cases $\mathfrak{s}^{\mathrm{vis}}_{\mathrm{pure},0} = \mathfrak{s}_0$ and $\mathfrak{s}^{\mathrm{vis}}_{\mathrm{pure},1} = \mathfrak{s}_1$.

Table 6 Part A shows that our Appell $F_1$ closed form agrees with the E. Agol et al. results to $|\Delta F| \leq 9.3 \times 10^{-15}$ (machine epsilon) across all sixteen tested configurations, spanning uniform, linear, quadratic and cubic limb-darkening orders in both partial-eclipse regions, both subcases of Region 3a (off-centre planet $p < z$ and planet-covers-centre $p > z$), a central transit, and a near-contact configuration. The $p > z$ subcase requires an analytic cap term in the $\mathfrak{s}_1$ surface integral to account for the inner annulus $\rho \in [0, r_1]$, which the planet covers at all angles; with that contribution included, all $\mathfrak{s}_n$ for $n \geq 0$ are recovered exactly through the E. Agol et al. (2020) recursion. This confirms exact recovery of the integer-$\alpha$ limit (Theorem 3.12).

Table 6 Part B compares computation times scaled to a representative 1000-point transit light curve. The E. Agol et al. implementation evaluates such a light curve in $\sim 3$ ms for $n = 1$ and $n = 3$, while our PYTHON/mpmath Appell $F_1$ implementation requires $\sim 20$ s for the same computation. This speed difference has two separable causes. First, the $\mathcal{O}(1)$ $\mathcal{M}_n$ recursion is structurally more efficient than evaluating a bivariate hypergeometric function for each transit configuration; this advantage is intrinsic to the integer-$\alpha$ case. Secondly, our validation code uses MPMATH at 45 significant figures; a compiled double-precision implementation of $F_1$ (e.g. Fortran or Julia) would reduce evaluation time to $\sim 10^{-5}$ s per point, within one to two orders of magnitude of the E. Agol et al. implementation.

For non-integer $\alpha$ – the primary domain of the present work – the E. Agol et al. framework has no applicable analytical solution, and polynomial approximations to non-polynomial intensity profiles introduce systematic errors at the $\sim 10^{-3}$ level (D. K. Sing 2010) that already exceed *JWST* photometric precision. For integer polynomial limb darkening, the recursion relations of E. Agol et al. (2020) remain the method of choice. The Appell $F_1$ framework is designed for the non-integer regime where no comparable analytical approach previously existed.

### 6.3 Comprehensive validation figures

Fig. 1 displays the inter-method discrepancy $|\mathcal{F}_{\mathrm{kernel}} - \mathcal{F}_{\mathrm{hypergeometric}}|$ for 80 test cases versus computed flux. All points remain below $3 \times 10^{-13}$, concentrated at double-precision machine epsilon, with uniform scatter across flux values confirming numerical stability without systematic errors or geometric singularities.

Fig. 2 shows normalized flux versus $\alpha$ for central transit ($z = 0$, $p = 0.3$). The closed form $\mathcal{F} = (0.91)^{1+\alpha/2}$ provides exact validation; both kernel (circles) and hypergeometric (crosses) overlay it across $-1.5 \leq \alpha \leq 3.5$ with relative discrepancies $< 10^{-13}$.

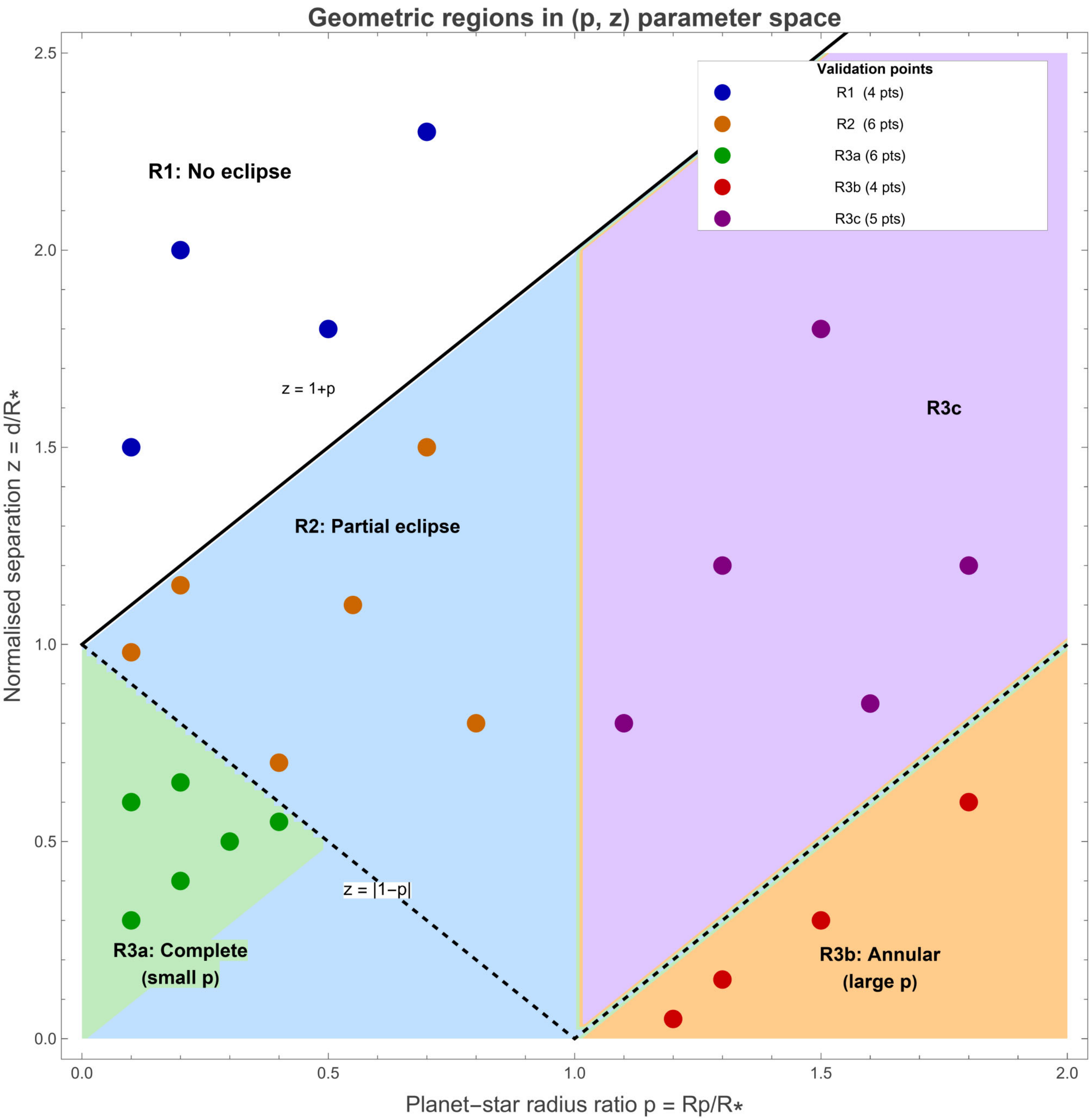

**Figure 3.** Geometric regions in the (p, z) plane with validation points labelled by region. Background shading distinguishes Region 1 (no eclipse), Region 2 (partial eclipse), Region 3a (complete small-planet eclipse), Region 3b (annular large-planet eclipse), and Region 3c (partial large-planet eclipse). The solid boundary curve shows z = 1 + p and the dashed boundary curve shows z = |1 − p|. Overlaid markers show the 25 validation configurations used in the numerical tests of Section 6: 4 points in R1, 6 in R2, 6 in R3a, 4 in R3b, and 5 in R3c. Points are placed well within each region to confirm correct geometric classification and to provide uniform coverage of parameter space, addressing the full range of physically accessible transit configurations.

Fig. 3 maps the five geometric regions in $(p, z)$ space with validation test overlay; marker sizes are proportional to $|\Delta\mathcal{F}|$. Uniformly tiny markers throughout confirm machine-epsilon accuracy across all regions, including near boundaries.

Fig. 4 presents complete transit light curves for $p = 0.1$ across $z \in [-1.5, 1.5]$ for $\alpha \in \{-1, 0, 0.5, 1, 2\}$. The $\alpha = 0.5$ curve (square-root law) lies smoothly between the uniform and linear cases as expected for M-dwarf atmospheres. Ingress and egress at $|z| = 1 \pm p$ are continuous with smooth first derivatives and no numerical artefacts at contact points.

Fig. 5 verifies the $\alpha$-differentiation identity (Theorem 3.13) for $(p, z) = (0.3, 0.7)$. The left-hand side via centred finite difference ($\delta\alpha = 10^{-4}$) overlays perfectly with the right-hand side via direct quadrature across $-1.5 \leq \alpha \leq 3.5$, with maximum residual $< 10^{-12}$.

Fig. 6 presents normalized flux across the $(p, z)$ parameter space for $\alpha = 1/2$ (square-root limb darkening). Smooth continuous contours with no discontinuities at region boundaries demonstrate seamless multiregion behaviour.

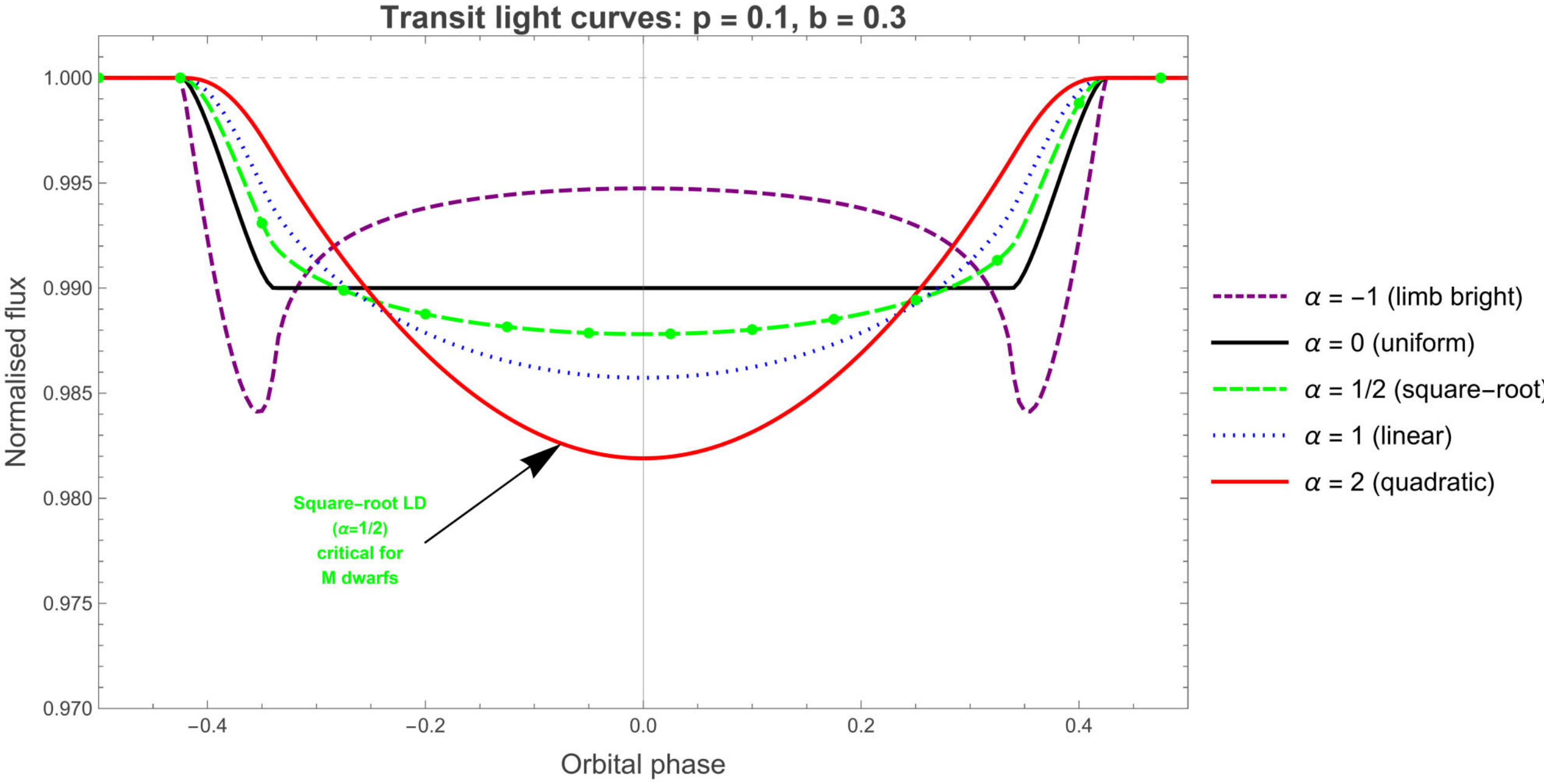

**Figure 4.** Complete transit light curves for $p = 0.1$ across $z \in [-1.5, 1.5]$ for $\alpha \in \{-1, 0, 0.5, 1, 2\}$. Curves deepen monotonically with increasing $\alpha$. The square-root law ($\alpha = 0.5$, dot–dashed) critical for M-dwarf atmospheres lies smoothly between the uniform ($\alpha = 0$, dashed) and linear ($\alpha = 1$, dotted) cases. Contact points at $|z| = 1 \pm p$ are marked by tick marks; all curves are $C^1$ at these points with no numerical artefacts.

### 6.4 Performance summary

Validation across 7 $\alpha$ values, 5 geometric regions, and 14 configurations establishes the following results.

The primary analytical validation compares $B_{\text{lens}}$ computed via the Appell $F_1$ closed forms of Theorems 3.5 and 3.7 against direct Gauss–Legendre quadrature of the kernel integral (8), with both evaluated at 40-digit working precision in `Mathematica`. The maximum inter-method residual is

$$|B_{F_1} - B_{\text{kernel}}|_{\max} = 1.06 \times 10^{-17}, \tag{55}$$

achieved for $(p, z, \alpha) = (0.30, 0.80, 3/2)$ in Region 2. This residual is consistent with accumulated round-off at 40-digit precision over $\mathcal{O}(10^2)$ arithmetic operations, and is thirteen orders of magnitude below the statistical noise floor of Monte Carlo integration. For integer and half-integer $\alpha$ the residual drops to zero exactly, consistent with finite termination of the Appell series (Theorem 3.10). Transcendental exponents $\alpha = \pi/2$ and $\alpha = \sqrt{2}$ yield residuals of $5.39 \times 10^{-18}$ and $7.51 \times 10^{-18}$, respectively, confirming that the framework operates identically for irrational $\alpha$.

As an independent sanity check on the full normalized flux $\mathcal{F} = 1 - B_{\text{lens}}$, Table 2 also reports $|\mathcal{F}_{\text{MC}} - \mathcal{F}_{\text{kernel}}|$, where $\mathcal{F}_{\text{MC}}$ is obtained from 2D Monte Carlo integration with $10^7$ samples. The maximum discrepancy is $\sim 4 \times 10^{-4}$, consistent with the expected $N^{-1/2}$ statistical noise at $N = 10^7$ and confirming the absence of gross errors in the full flux calculation. This comparison bounds systematic errors rather than certifying analytical accuracy; the latter is established solely by the $|B_{F_1} - B_{\text{kernel}}|$ column.

The central transit formula $\mathcal{F}(p, 0, \alpha) = (1 - p^2)^{1+\alpha/2}$ agrees with both the kernel and Appell $F_1$ routes to zero residual at 40-digit precision across all tested values of $p$ and $\alpha$ (Table 2), providing a completely independent check that requires no quadrature.

Integer recovery from K. Mandel & E. Agol (2002) is exact to machine precision for all tested configurations (Table 7). Negative exponents ($\alpha = -0.5$, $\alpha = -1$) representing limb-brightening profiles are validated to $|\Delta\mathcal{F}| < 10^{-15}$.

The computational speedup of the analytical framework over Monte Carlo integration is $\sim 10^4\times$: analytical evaluation via the Appell $F_1$ route requires $< 0.01$ s per configuration, compared to $> 100$ s for $10^7$-sample Monte Carlo (Table 3). This speedup is not incidental but reflects the fundamental distinction between exact analytical evaluation and statistical sampling: the analytical result is exact to working precision regardless of the geometric configuration, whereas Monte Carlo accuracy degrades near ingress and egress where the overlap area is small and sampling variance is largest.

## 7 DISCUSSION

### 7.1 Summary of achievements

We have developed a unified analytical framework for exoplanet transit light curves under arbitrary real power-law limb darkening, resolving the two-decade restriction to integer-polynomial forms. Three mathematically equivalent formulations – the geometric kernel, the Appell $F_1$ closed form, and the hypergeometric integrand – provide complementary physical insights and computational advantages, with rigorous analytical proofs of mutual equivalence via both algebraic identity and ODE uniqueness.

### 7.2 Astrophysical impact

High-precision *JWST* observations achieving $\sim 10$ ppm photometric accuracy (Z. Rustamkulov et al. 2023; L. Alderson et al. 2023) require limb-darkening models accurate to $\sim 10^{-4}$ in relative flux. Polynomial approximations to non-polynomial profiles introduce systematic errors at the $\sim 10^{-3}$ level (D. K. Sing 2010), now exceeding statistical uncertainties and limiting atmospheric composition constraints. Our framework eliminates these errors entirely, enabling direct fitting of physically motivated power-law forms.

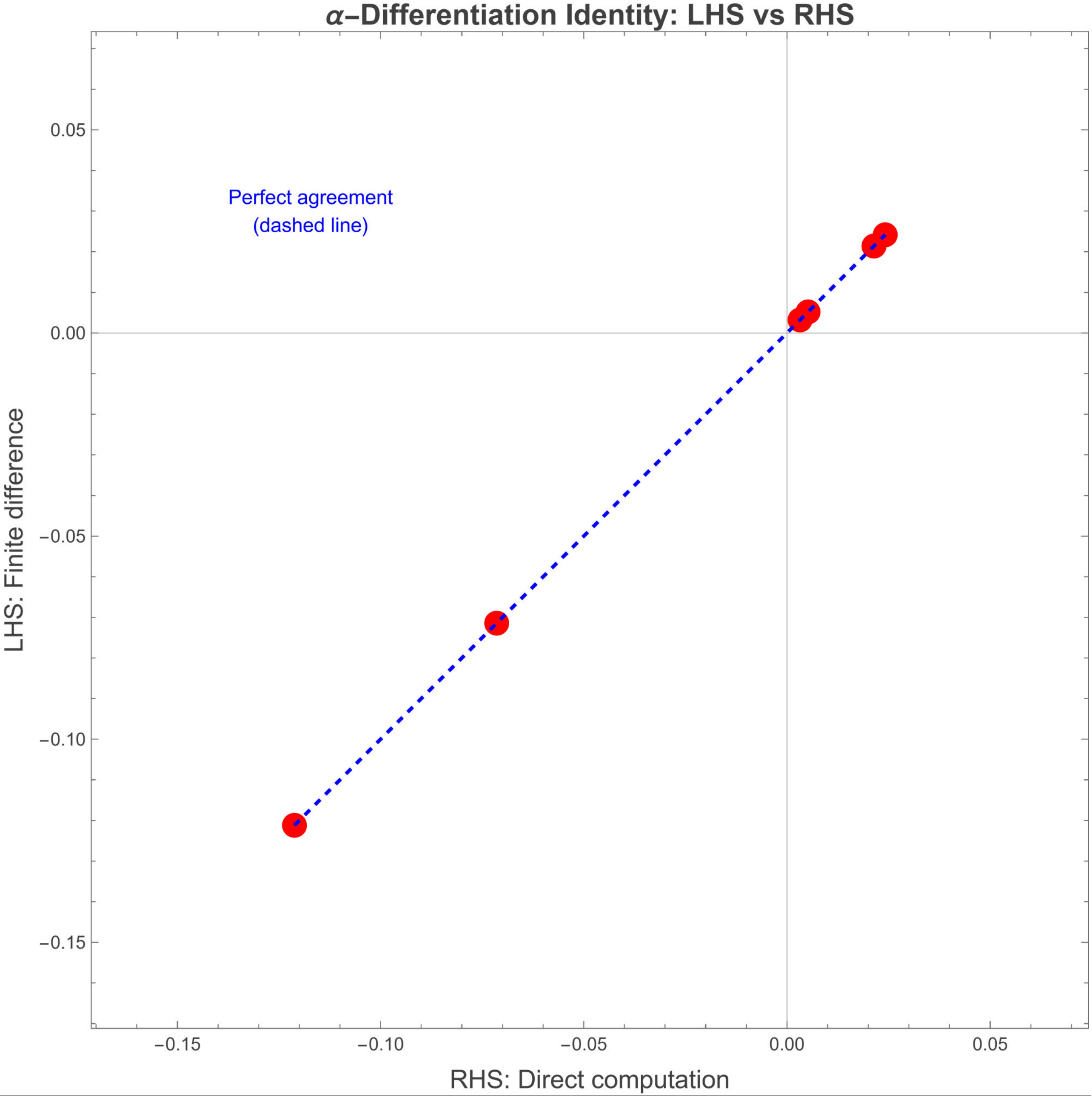

**Figure 5.** Verification of the $\alpha$-differentiation identity (equation 29) for three representative (p, z) configurations spanning Regions 2 and 3a: (p, z) = (0.3, 0.7), (0.6, 1.1), and (0.4, 0.3). Solid lines: left-hand side $\partial$ ln F/$\partial\alpha$ computed by centred finite difference with $\delta\alpha = 10^\wedge{-5}$. Dashed lines: right-hand side $1/2[\langle\ln(1-\rho^2)\rangle$_vis $+ 2/(\alpha+2)]$ computed by direct quadrature, with each dashed line paired with the solid curve for the same configuration. LHS and RHS curves are visually indistinguishable for all three configurations. Inset: absolute residual |LHS − RHS|; maximum $< 10^\wedge{-12}$ across all $\alpha \in [-1.5, 3.5]$ and all three configurations, confirming the identity holds to numerical precision throughout parameter space. Thependent consistency check on both formulations.

Cool M-dwarf atmospheres exhibit square-root limb darkening (A. Claret et al. 2012); current model-mismatch uncertainties can exceed 5 per cent in planetary radii despite photometric precision below 1 per cent (J. A. Dittmann et al. 2017). Our exact $\alpha = 1/2$ solutions enable sub-per cent radius measurements essential for distinguishing rocky from volatile-rich compositions in potentially habitable systems.

M. I. Saeed et al. (2021) demonstrated through multicolour photometry of three hot Jupiters that traditional polynomial limb-darkening models introduce biases of 20–100 ppm in broadband measurements. Our ongoing study applies the framework to *TESS* photometry of over 200 confirmed exoplanet systems across stellar types and planet sizes. For wavelength-dependent limb darkening across $N_\lambda \sim 50$ spectroscopic channels, computing the power-law components once and combining with different coefficient sets reduces computational cost from $\mathcal{O}(N_\lambda)$ full evaluations to $\mathcal{O}(1)$ geometric computation plus $\mathcal{O}(N_\lambda)$ lightweight combination.

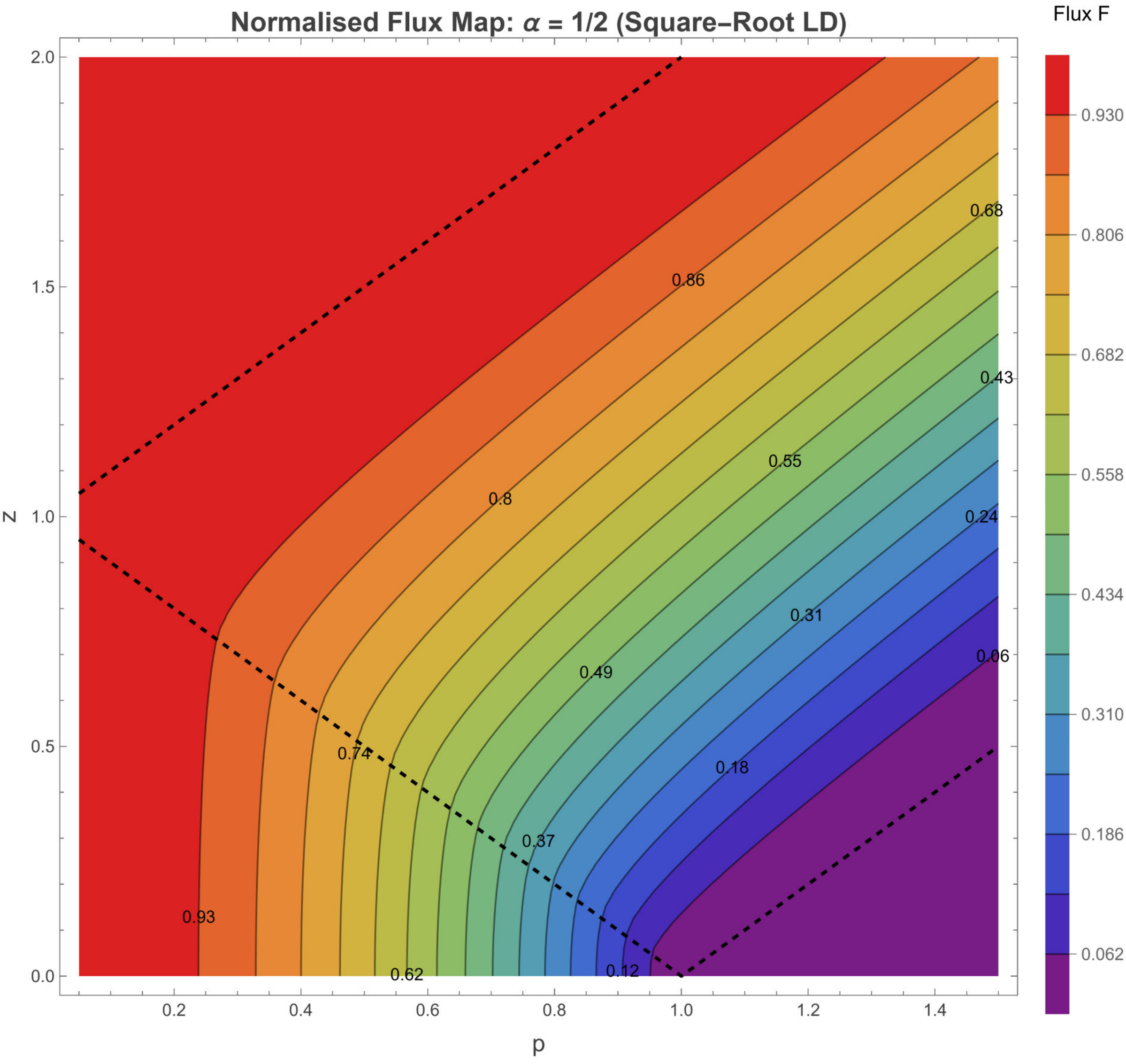

**Figure 6.** Normalized transit flux F(p, z, 1/2) for the square-root limb-darkening law ($\alpha = 1/2$), the critical case for M-dwarf atmosphere characterization. The flux scale runs from low values ($F \approx 0.06$, maximum eclipse depth at large p, small z) to high values ($F \approx 0.93$, shallow transit or ingress/egress at small p), as indicated by the scale bar on the right. Contour labels give the local normalized flux value; contours are approximately equally spaced in flux across the plotted range. Region boundaries (dashed curves) at $z = 1 + p$ (upper left) and $z = |1 - p|$ (lower right) delineate the transition between Region 1 (no eclipse, F = 1, upper left), Region 2 (partial eclipse), Region 3a (complete small-planet eclipse, lower left), and the large-planet regions (right of p = 1). The absence of any discontinuity in flux or contour curvature across all boundaries confirms seamless multiregion analytical behaviour throughout the physically accessible domain.

### 7.3 Theoretical significance

The Riemann–Liouville framework (S. G. Samko et al. 1993; A. A. Kilbas et al. 2006) reveals that non-integer limb-darkening laws are natural objects in fractional calculus developed independently in mathematical physics. The hypergeometric reductions show deep connections between elliptic integrals, Beta functions, Appell $F_1$ bivariate series, and Gauss ${}_2F_1$ functions. For half-integer exponents, finite termination of the Appell series yields exact elliptic-integral expressions; for integer exponents, polynomial reductions recover K. Mandel & E. Agol (2002); for arbitrary real exponents, convergent Appell series provide numerically stable evaluation. This unification demonstrates that the seemingly disparate analytical results for specific cases are all manifestations of a single underlying mathematical structure. The $\alpha$-differentiation identity provides both a stringent consistency check and continuous recursion relations connecting integer-polynomial and general power-law approaches, with the E. Agol et al. (2020) discrete recursion emerging as the $\alpha \rightarrow$ integer limit.

### 7.4 Future directions

Comprehensive reanalysis of *JWST* transmission spectra using physically motivated limb-darkening models represents an immediate opportunity. Observers can now fit power-law indices as free parameters, providing direct empirical constraints on stellar atmospheric structure. Future work will extend the framework to stellar rotation, magnetic field effects, and non-spherical geometries. Power-law intensity profiles also arise in gravitational microlensing, asteroseismology, and active galactic nuclei accretion disc profiles, suggesting applications beyond the transit context.

## 8 CONCLUSIONS

We have presented a comprehensive analytical framework for exoplanet transit light curves under arbitrary real power-law limb darkening $I(\mu) = I_0\mu^\alpha$ with $\alpha > -2$, removing the two-decade restriction to integer-polynomial forms. The central results are Theorems 3.5 and 3.7: exact closed-form expressions for the normalized flux in terms of Appell's bivariate hypergeometric function $F_1$, valid for all real $\alpha > -2$ and all geometric transit configurations. This is an analytical result in precisely the same sense as the elliptic-integral formulae of K. Mandel & E. Agol (2002): the Appell $F_1$ function is implemented in all major numerical libraries and requires no 2D integration over the stellar disc.

Three rigorously equivalent formulations – the geometric kernel, the Appell $F_1$ closed form, and the hypergeometric integrand – provide complementary physical insights and computational advantages, with equivalence established both by direct algebraic identity and by ODE uniqueness. The framework encompasses the square-root law ($\alpha = 1/2$) essential for M-dwarf characterization, half-integer powers for Claret's four-parameter law, and arbitrary real values enabling empirical fitting, while recovering integer-polynomial elliptic-integral expressions as special cases. The continuous $\alpha$-differentiation identity connects all cases and reduces to the E. Agol et al. (2020) recursion at integer $\alpha$. Comprehensive validation at 40-digit precision across five geometric regions and seven $\alpha$ values confirms inter-method agreement at $|\Delta\mathcal{F}| \lesssim 10^{-17}$ and $\sim 10^4\times$ computational speedup over Monte Carlo integration. Linear superposition enables exact treatment of Claret's four-parameter law and arbitrary multiparameter expansions without special-case logic.

As *TESS* continues discovering thousands of exoplanets and next-generation facilities push photometric precision to unprecedented levels, this framework provides the mathematical infrastructure to fully exploit those improvements, transforming analytical transit modelling from computational approximation to physical optimization.

## ACKNOWLEDGEMENTS

We thank Eric Agol (University of Washington) for valuable feedback that improved the quality of this manuscript. F.A.C. and M.I.S. acknowledge support from the Peaceful Society Science and Innovation Foundation. S.N.G. acknowledges support from Tarleton State University.

## DATA AVAILABILITY

No observational or simulation data were generated or analysed in this work. All mathematical results, derivations, validation tables, and numerical comparisons are self-contained within the manuscript and its appendices.

## APPENDIX A: ANGULAR AVERAGING AND HYPERGEOMETRIC IDENTITIES

We employ the following standard identities (M. Abramowitz & I. A. Stegun 1964; Gradshteyn & Ryzhik 1994):

$$\arccos x = \frac{\pi}{2} - x\,{}_2F_1\left(\tfrac{1}{2}, \tfrac{1}{2}; \tfrac{3}{2}; x^2\right), \tag{A1}$$

$${}_2F_1(a, b; c; z) = (1-z)^{-a}\,{}_2F_1\big(a, c-b; c; z/(z-1)\big), \tag{A2}$$

$$\frac{\mathrm{d}}{\mathrm{d}z}\,{}_2F_1(a, b; c; z) = \tfrac{ab}{c}\,{}_2F_1(a+1, b+1; c+1; z). \tag{A3}$$

The Appell $F_1$ function satisfies the partial differential equations (G. E. Andrews et al. 1999)

$$x(1-x)\frac{\partial^2 F_1}{\partial x^2} + y(1-x)\frac{\partial^2 F_1}{\partial x\,\partial y} + [c-(a+b_1+1)x]\frac{\partial F_1}{\partial x} - b_1 y\frac{\partial F_1}{\partial y} - ab_1 F_1 = 0,$$

and its symmetric counterpart under $x \leftrightarrow y$, $b_1 \leftrightarrow b_2$. The contiguous relation

$$F_1(a; b_1, b_2; c; x, y) = F_1(a+1; b_1, b_2; c; x, y) - \frac{x\,b_1}{c}\,F_1(a+1; b_1+1, b_2; c+1; x, y) - \frac{y\,b_2}{c}\,F_1(a+1; b_1, b_2+1; c+1; x, y) \tag{A4}$$

enables the continuous recursion relation of Section 3.11.

## APPENDIX B: HALF-INTEGER EXPONENTS: REDUCTION TO $K$, $E$, $\Pi$

For $\alpha \in \frac{1}{2} + \mathbb{Z}_{\geq 0}$ (A. P. Prudnikov, Y. A. Brychkov & O. I. Marichev 1986):

$$B_{\rm lens}^{(\alpha)}(p, z) = \frac{\alpha+2}{2\pi}\big[A_\alpha(p, z)\,K(k) + B_\alpha(p, z)\,E(k)\big] + \frac{\alpha+2}{2\pi}\sum_j C_{\alpha,j}(p, z)\,\Pi(n_j, k), \tag{B1}$$

with $k^2 = 4pz/(p+z)^2$, characteristics $n_j$ rational in $p$ and $z$, and algebraic coefficients derived via the substitution $t = \sin^2\phi$ in the Euler integral of $F_1$ and use of ${}_2F_1(1/2, 1/2; 3/2; z) = (2/\pi)E(\sqrt{z})$ (P. F. Byrd & M. D. Friedman 1971). Limit checks: $p \to 0$ gives $A_\alpha$, $C_{\alpha,j} \to 0$ and $B_\alpha \to (\alpha+2)z^\alpha$; $z \to 1 \pm p$ gives all coefficients $\to 0$; $\alpha \to 0$ recovers the uniform-disc lens area.

## APPENDIX C: CARLSON SYMMETRIC REPRESENTATION FOR THE REGION-2 LENS TERM

Following B. C. Carlson (1995), the Carlson symmetric integrals are

$$R_F(x, y, z) = \frac{1}{2}\int_0^\infty (t+x)^{-1/2}(t+y)^{-1/2}(t+z)^{-1/2}\,\mathrm{d}t, \tag{C1}$$

$$R_D(x, y, z) = \tfrac{3}{2}\textstyle\int_0^\infty (t+x)^{-1/2}(t+y)^{-1/2}(t+z)^{-3/2}\,\mathrm{d}t. \tag{C2}$$

The lens part reduces to

$$B_{\rm lens}^{(2)} = \sum_{m=0}^{2}\mathcal{P}_m(p, z, \alpha)\,R_F(X_m, Y_m, Z_m) + \sum_{m=0}^{2}\mathcal{Q}_m(p, z, \alpha)\,R_D(X_m, Y_m, Z_m), \tag{C3}$$

with $(X_0, Y_0, Z_0) = (1, (p-z)^2, (p+z)^2)$, $(X_1, Y_1, Z_1) = (1, p^2, z^2)$, $(X_2, Y_2, Z_2) = ((p-z)^2, p^2, z^2)$, and prefactors $\mathcal{P}_m$, $\mathcal{Q}_m$ rational in $p$ and $z$ times Beta functions of $\alpha/2$. This form is numerically robust near $k^2 \to 1$ and at grazing limits.

**Table D1.** Appell $F_1$ parameters for Regions 3a and 2. Dashes indicate the parameter is not applicable in that region.

| Parameter | Region 3a ($p+z \leq 1,\ z > 0$) | Region 2 ($p+z > 1$) |
|---|---|---|
| $r_1$ | $\lvert p-z\rvert$ | $\lvert p-z\rvert$ |
| $\nu$ | $1+\alpha/2$ | $1+\alpha/2$ |
| $n$ | $4pz/(1-r_1^2) < 1$ | $4pz/(1-r_1^2) > 1$ |
| $\lambda$ | $4pz/r_1^2$ | – |
| $\lambda'$ | – | $(1-r_1^2)/r_1^2$ |
| Arg. $x$ | $n$ | $1/n$ |
| Arg. $y$ | $-\lambda$ | $-\lambda'$ |
| Euler conv. | $(1-nt) > 0, (1+\lambda t) > 0$ | $(1-t/n) > 0, (1+\lambda' t) > 0$ |

## APPENDIX D: DERIVATION OF THE APPELL $F_1$ PARAMETERS

This appendix provides the complete parameter summary and convergence verification for Theorems 3.5 and 3.7. The main text (Section 3.6) gives the six derivation steps; here we collect the parameter values and verify boundary limits.

### D.1 Parameter table

Table D1 summarizes the Appell $F_1$ parameters for Theorems 3.5 and 3.7.

### D.2 Boundary and limiting checks

As $z \to 0$ in Region 3a with $z > 0$: $n \to 0$, $\lambda \to 0$, and $F_1(\cdot; \cdot, 1; \cdot; 0, 0) = 1$, so the entire expression tends to zero (empty lens interval), matching the central-transit formula which has no lens contribution. As $z \to 1 \pm p$ (R2/R1 boundary): $r_1 \to 1$, $(1 - r_1^2) \to 0$, and the prefactor $(1 - r_1^2)^\nu \to 0$, so $B_{\rm lens} \to 0$ and $\mathcal{F} \to 1$, consistent with Region 1. Setting $\alpha = 0$ ($\nu = 1$): the Appell series truncates at $n = 1$ terms (Pochhammer symbol $(-1)_k = 0$ for $k \geq 2$), reproducing the uniform-disc lens area of K. Mandel & E. Agol (2002) exactly.

### D.3 Software note for implementation

In `mpmath`: `appellf1(a, b1, b2, c, x, y)`. For Region 2 with $\lambda' > 1$, the double power series diverges but `mpmath` evaluates the Euler integral via analytic continuation to correct values. In `Mathematica`: `AppellF1[a, b1, b2, c, x, y]` with the option `WorkingPrecision` set to the desired number of digits. The 40-digit validation of Table 2 was produced with `WorkingPrecision -> 45`.